\documentclass[aps,prb,twocolumn,groupedaddress]{revtex4}
\usepackage{psfig,graphics}
\usepackage{color,mycolors}
\usepackage{amssymb}
\usepackage{bookmath}
\begin{document}
%~~~~~~~~~~~~~~~~~~~~~~~~~~~~~~~~~~~~~~~~~~~~~~~~~
\preprint{NU preprint: cmt116}
\title{Thermodynamic properties of thin films of superfluid \Hea}
%\homepage[eprint:]{http://snowmass.phys.nwu.edu/~sauls/Eprint/cmt110}
\author{A. B. Vorontsov}
%\email[contact:]{anton@snowmass.phys.nwu.edu}
\author{J. A. Sauls}
\affiliation{Department of Physics and Astronomy,
             Northwestern University, Evanston, IL 60208}
\date{\today}

\begin{abstract}
The pairing correlations in superfluid \He\ are strongly modified by quasiparticle
scattering off a surface or an interface. We present theoretical results and
predictions for the order parameter, the quasiparticle excitation spectrum and
the free energy for thin films of superfluid \He. Both specular and diffuse
scattering by a substrate are considered, while the free surface is assumed to be
a perfectly reflecting specular boundary. The results are based on
self-consistent calculations of the order parameter and quasiparticle
excitation spectrum at zero pressure. We obtain new results for the phase
diagram, free energy, entropy and specific heat of thin films of superfluid \He.
\end{abstract}
\maketitle
%~~~~~~~~~~~~~~~~~~~~~~~~~~~~~~~~~~~~~~~~~~~~~~~~~

\section{\label{sec:intro} Introduction}

The superfluid phases of bulk \He\ are condensates of p-wave,
spin-triplet Cooper pairs. The order parameters describing the A-
and B-phases spontaneously break orbital- and spin-rotational
symmetries, as well as discrete symmetries of space- and
time-inversion, of the normal Fermi-liquid phase of \He.\cite{vol90}
Pairing transitions of this type are called ``unconventional'', and such
systems exhibit novel phenomena associated with the spontaneously
broken symmetries of the order parameter. A generic feature of
superfluids with an unconventional order parameter is their
sensitivity to scattering by impurities, defects and boundaries.
Scattering of quasiparticles by these objects leads to the
suppression of the order parameter (pairbreaking), suppression of
the superfluid transition as well as more subtle physical effects
associated with the interplay of scattering and particle-hole
coherence in the superfluid state. These effects persist over
several coherence lengths, which for \He\ is
$\xi_0=\hbar v_f/2\pi k_B T_c\simeq 10^2-10^3$ {\AA}.
When \He\ is placed in a container or confined
to a geometry with dimensions of order the coherence length scale
the effects of the boundaries on the superfluid extend to all
parts of the liquid, and there is in this sense no ``bulk'' phase.

Superfluidity of \He\ films was first reported in
1985.\cite{sac85} Basic properties of superfluid films, such
as the transition temperature, critical current, superfluid
density, have been measured in several
laboratories.\cite{dav88,dau88,ste94} Evidence of a presumed A-B
phase transition has been reported by several groups, and a phase
diagram has been constructed over a limited range of temperatures
and film thickness.\cite{fre88,xu90,kaw98} Experimental
observation of third sound in \He\ films also shows anomalies in
the mode spectrum as a function of temperature and film thickness
which cannot be accounted for within the hydrodynamic theory
applicable to superfluid \Hefour.\cite{sch98} There are many
open questions about the nature of the superfluid phases of \He\
in restricted geometries, including the identification of the
superfluid order parameter and whether or not additional phases may
be stabilized depending on the geometry and surface structure of
the confining geometry.

Theoretical models and calculations of the surface structure and
excitation spectrum of superfluid \He\ films are important for
understanding both thermodynamic and nonequilibrium properties of
superfluid films. Detailed comparison between theory and
experiment for the temperature dependence of the heat capacity and
entropy, allows us to determine the density of states and low lying
excitations of the film.  The heat capacity and entropy
also enter the hydrodynamic equations that describe
damping of low frequency collective modes of superfluid films.

In this paper we report a theoretical study of superfluid \He\ in
films with thickness ranging from $D\sim 1\,-\,15\,\xi_0$. This is
a system similar to \He\ in a slab, except the \He\ is in general
confined between different interfaces. Essential to any theory of
superfluid \He\ in confined geometry are the boundary conditions
that describe the effects of surface scattering on the pairing
correlations, the quasiparticle spectrum and quasiparticle
distribution functions in the case of nonequilibrium properties.

Early theoretical investigations focused on pairbreaking of the
order parameter near a wall in the Ginzburg-Landau (GL) regime, and
the implications of the pair-breaking effect on the boundary
conditions for the hydrodynamic variables describing the A-phase,
in particular the $\vell$ vector.\cite{amb75} These calculations,
as well as calculations of the suppression of the transition
temperature for superfluid \He\ in confined geometry,\cite{kja87}
were based on deGennes' formulation of inhomogeneous superfluidity
in terms of semiclassical correlation functions and a heuristic
model of surface roughness which interpolated between
specular and diffuse scattering of quasiparticles.
More recent analyses based on the GL theory for the phase diagram
and dynamical properties of \He\ in slabs and cylindrical pores are
described in Refs. \onlinecite{li88,fet88}.

Extensions of surface pairbreaking calculations beyond the GL
limit require a more detailed theoretical formulation of
inhomogeneous states of superfluid \He. The most powerful theory
of superfluid \He\ is based on the quasiclassical transport
equations.\cite{eil68,lar68,eli71} This theory is the natural
extension of Landau's theory of normal Fermi liquids to include
BCS pairing correlations. The quasiclassical theory is applicable
to a broad range of phenomena and nonequilibrium states of
inhomogeneous \He. The central objects of the quasiclassical
theory are the propagators that describe both the quasiparticle
excitations of the condensed phases and the correlated pairs that
form the condensate. Theoretical calculations based on the
quasiclassical theory for the surface order parameter and
excitation spectrum near a wall or interface require boundary
conditions for the quasiclassical propagators, generally
formulated from scattering theory and a specific model for the
surface or interface.

Boundary conditions describing reflection from an atomically rough
surface were developed by Buchholtz and Rainer\cite{buc79} and
implemented in the form of the randomly rippled wall (RRW)
approximation for \Heb.\cite{buc86,buc91} Alternative formulations of
diffuse boundary conditions were implemented by Zhang et
al.\cite{zha87} based on scattering from a thin layer of atomic-size
impurities coating an otherwise smooth surface, and by Thuneberg, et
al.\cite{thu92b} based on scattering from a distribution of ``randomly
oriented mirror'' (ROM) surfaces. Boundary conditions describing rough surfaces
which are neither perfectly specular nor fully diffuse were developed
by Nagato et al.\cite{nag96} in terms of a random S-matrix. In the diffuse
scattering limit the results of Ref. \onlinecite{nag98} for the order
parameter suppression in \Heb\ agree well with those obtained by Zhang
et al.\cite{zha87} based on the TDL model.

We report a theoretical analysis of the structure, excitation
spectrum and thermodynamic properties of supefluid \He\ films.
The free surface is modelled as a specular surface, and the film
resides on a substrate that we assume is atomically rough. We
model the scattering of quasiparticles by the substrate by
introducing a thin layer of atomic-size impurities randomly
distributed on the surface (TDL model). The width, $d$, of the
impurity layer is assumed to be much less than the superfluid
coherence length, while the mean free path $l_{\text{imp}}$ for
quasiparticles propagating in the layer is much smaller than the
width $d$. Thus, quasiparticles are strongly scattered inside the
layer, but eventually scatter out of the layer at an angle
uncorrelated with the incident trajectory. The limit
$d/l_{\text{imp}}\to\infty$ as $d\to0$ describes a rough surface
in the diffuse scattering limit. The specfic formulation of the
impurity model for diffuse scattering that we use was
introduced by Ovchinnikov\cite{ovc69} for diffuse scattering from
atomically rough metallic interfaces, and was implemented for
superfluid \He\ by Kopnin.\cite{kop91}

We start from the quasiclassical theory of superfluidity with the
boundary conditions described above and calculate the equilibrium
properties of films of superfluid \He\ in the weak-coupling limit.
We calculate suppression of the order parameter in the film, the
quasiparticle density of states, superfluid free energy, entropy
and heat capacity. We are especially interested in films of
thickness $D \lesssim D_{\text{\tiny AB}}$, where
$D_{\text{\tiny AB}}\approx 9\xi_0$ marks the transition
from the B-like phase for thicker films to the A-like phase for
thinner films. It is in this region that most film
experiments have been done. We find that because of diffuse scattering
at one of the interfaces, a band of subgap states is formed and
these states exist throughout the film for sufficiently
thin films. The thermodynamic properties of the A-phase are
changed significantly by these gapless excitations.

The main sections of this paper are organized as follows. In
section \ref{sec:model} we describe the theoretical formulation of
the quasiclassical transport equations and the boundary conditions
that we use in our calculations. Results for the thin film phase
diagram are summarized in section \ref{sec:phases}. In section
\ref{sec:dos} we discuss the excitation spectrum for both specular
and diffuse scattering, while in section \ref{sec:fe} we present
the results for the free energy, entropy and heat capacity of thin
films. More technical aspects of computing the free
energy and implementing the diffuse boundary condition in the
Riccati formulation of the transport equations are included in
appendices.

\section{\label{sec:model} Theoretical Model}

\begin{figure}[ht]
\centerline{\psfig{figure=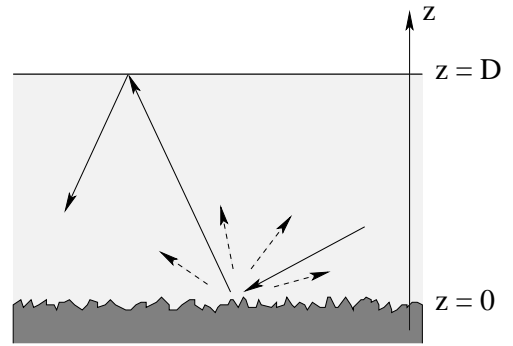,height=4.5cm}}
\caption{\label{fig:model} A thin film of \He\ on an
atomically rough substrate (at $z=0$) with a free surface ($z=D$) which is specular. }
\end{figure}

We assume that the \He\ film is on a substrate that is
atomically rough on a scale much shorter than
coherence length, $\xi_0$. The film has a well-defined
thickness, $D$, that is larger than the atomic scale.
The free surface of the film is assumed to be
atomically smooth, {\it c.f.} Fig. \ref{fig:model}.
We assume there is negligible evaporation and vapor above the film.
Thus, the liquid in the film is essentially under zero pressure.
In this model \He\ quasiparticles are specularly
reflected at the free surface. We also assume that
film is invariant under translations and rotations in
plane of the film ($xy$-plane),\footnote{Rotational
symmetry is broken so to be precise we consider
superfluid films in which rotational symmetry
about the $\hat\vz$-axis is preserved
under joint rotation of the spin and orbital degrees
of freedom up to an overall gauge transformation.}
and thus the physical properties of the film depend
only on $z$-coordinate, which is normal to the
substrate and the free surface.

%~~~~~~~~~~~~~~~~~~~~~~~~~~~~~~~~~~~~~~~~~~~~~~~~~~~~~~~~~~~~~~~~~~~~~~~~~~~~~~~~
%~~~~~~~~~~~~~~~~~~~~~~~~~~~~~~~~~~~~~~~~~~~~~~~~~~~~~~~~~~~~~~~~~~~~~~~~~~~~~~~

The calculations we report were carried out using the quasiclassical theory
for superfluid \He,\cite{ser83} supplemented by boundary conditions for
surface scattering at the vapor-liquid interface and the film substrate.
The central object of the quasiclassical theory is the
propagator, $\whg(\hat{\vp}, \vR; \varepsilon_m)$, which is a $4\times 4$
Nambu matrix - denoted by a widehat - in the combined particle-hole and spin spaces,
and is defined in terms of an integration of the full Nambu propagator,
%~~~~~~~~~~~~~~~~~~~~~~~~~~~~~~~~~~~~~~~~~~~~~~~~~~~~~~~~~~~~~~~~~~~~~~~~~~~~~~~~~~~~~~~~~~~~
\ber
\whG(\vp,\vR;\varepsilon_m)&=&\int_0^{\beta}\,d\tau\,e^{i\varepsilon_m\tau}
                              \int\,d^3r\,e^{-i\vp\cdot\vr}\times\\
&-&\langle\T_{\tau}\Psi(\vR+\vr/2,\tau)\bar{\Psi}(\vR-\vr/2,0)\rangle
\,,\nonumber
\eer
%~~~~~~~~~~~~~~~~~~~~~~~~~~~~~~~~~~~~~~~~~~~~~~~~~~~~~~~~~~~~~~~~~~~~~~~~~~~~~~~~~~~~~~~~~~~~
over a shell, $|v_f(p-p_f)|<\varepsilon_c\ll E_f$, in momentum space near the Fermi surface,
%~~~~~~~~~~~~~~~~~~~~~~~~~~~~~~~~~~~~~~~~~~~~~~~~~~~~~~~~~~~~~~~~~~~~~~~~~~~~~~~~~~~~~~~~~~~~
\be
\whg(\hat{\vp},\vR;\varepsilon_m)=\frac{1}{a}
\int_{-\varepsilon_c}^{+\varepsilon_c}
d\xi_{\vp}\,\widehat{\tau}_3\whG(\vp,\vR;\varepsilon_m)
\,.
\ee
%~~~~~~~~~~~~~~~~~~~~~~~~~~~~~~~~~~~~~~~~~~~~~~~~~~~~~~~~~~~~~~~~~~~~~~~~~~~~~~~~~~~~~~~~~~~~
The propagator is normalized by dividing by the weight of the quasiparticle pole in the
spectral function, $a$.
We use the Matsubara representation to calculate equilibrium properties; the
fermion Matsubara frequencies are $\varepsilon_m=(2m+1)\pi k_B T$.
The four-component Nambu field operators are defined in terms of the bare
fermion field operators by
$\Psi=(\psi_{\uparrow},\psi_{\downarrow},
                 \psi^{\dagger}_{\uparrow},\psi^{\dagger}_{\downarrow})$,
and $\bar{\Psi}(\vr,\tau)=\Psi^{\dagger}(\vr,-\tau)$.

For pure spin-triplet pairing the quasiclassical propagator $\whg$ may be
parametrized in particle-hole space by $2\times 2$ spin matrices
for the diagonal (quasiparticle) and off-diagonal (Cooper pair) propagators,
%~~~~~~~~~~~~~~~~~~~~~~~~~~~~~~~~~~~~~~~~~~~~~~~~~~~~~~~~~~~~~~~~~~~~~~~~~~~~
\be
\whg =
\left( \begin{array}{cc}
\hg & \hf \\
\ul{\hf} & \ul{\hg}
\end{array} \right)
=
\left( \begin{array}{cc}
g+\hat{\vsigma}\cdot\vg & (i\hat{\vsigma}\hat{\sigma}_y) \cdot \vf \\
(i\hat{\sigma}_y\hat{\vsigma})\cdot \ul{\vf} &
\ul{g} + \hat{\vsigma}^{\text{\tiny tr}}\cdot\ul{\vg}
%\ul{g} - \hat{\sigma}_y\hat{\vsigma}\hat{\sigma}_y\cdot\ul{\vg}
\end{array} \right)
\,.
\label{eq:gmt}
\ee
%~~~~~~~~~~~~~~~~~~~~~~~~~~~~~~~~~~~~~~~~~~~~~~~~~~~~~~~~~~~~~~~~~~~~~~~~~~~~
The $2\times 2$ spin matrices are denoted by ordinary hats, e.g. $\hg$. The spin vector,
$\hat{\vsigma}=\left(\hat{\sigma}_x,\hat{\sigma}_y,\hat{\sigma}_z\right)$, are the Pauli matrices.
One deviation from the matrix notation is $\hat{\vp}$, which denotes a unit vector in the
direction of the Fermi velocity, $\vv_f(\hat{\vp})= v_f\hat{\vp}$.
The components of the quasiclassical propagator are not all independent. The upper and lower
particle-hole components are related by symmetries that follow from
the fermion anticommutation relations,
%~~~~~~~~~~~~~~~~~~~~~~~~~~~~~~~~~~~~~~~~~~~~~~~~~~~~~~~~~~~~~~~~~~~~~~~~~~~~
\be
\begin{array}{l}
\displaystyle
\ul{\hf}(\hat{\vp}, \vR; \varepsilon_m) = \hf(-\hat{\vp}, \vR; \varepsilon_m)^* =
- \hf(\hat{\vp}, \vR; -\varepsilon_m)^\dag \,,\\
\displaystyle
\ul{\hg}(\hat{\vp}, \vR; \varepsilon_m) = \hg(-\hat{\vp}, \vR; \varepsilon_m)^* =
\hg(-\hat{\vp}, \vR; -\varepsilon_m)^{\text{tr}} \,,
\end{array}
\label{eq:gsm}
\ee
%~~~~~~~~~~~~~~~~~~~~~~~~~~~~~~~~~~~~~~~~~~~~~~~~~~~~~~~~~~~~~~~~~~~~~~~~~~~~
where $\hg^{\text{tr}}$ is the matrix transpose of $\hg$.

The quasiclassical transport equation that governs the evolution of the quasiclassical
propagator, $\whg(\hat{\vp}, \vR; \varepsilon_m)$, is\cite{eil68,ser83}
%~~~~~~~~~~~~~~~~~~~~~~~~~~~~~~~~~~~~~~~~~~~~~~~~~~~~~~~~~~~~~~~~~~~~~~~~~~~~
\bea
\big[i\varepsilon_m \widehat{\tau}_3 - \whDelta(\hat{\vp},\vR)
&,& \whg(\hat{\vp}, \vR; \varepsilon_m)\big] \nonumber \\
&+& i\vv_f(\hat{\vp})\cdot\grad\whg(\hat{\vp},\vR;\varepsilon_m) = 0 \,,
\label{eq:transport_equation}
\eer
%~~~~~~~~~~~~~~~~~~~~~~~~~~~~~~~~~~~~~~~~~~~~~~~~~~~~~~~~~~~~~~~~~~~~~~~~~~~~
with a constraint given by Eilenberger's normalization condition on $\whg$,
%~~~~~~~~~~~~~~~~~~~~~~~~~~~~~~~~~~~~~~~~~~~~~~~~~~~~~~~~~~~~~~~~~~~~~~~~~~~~
\be
\whg(\hat{\vp}, \vR; \varepsilon_m)^2 = -\pi^2 \widehat{1} \,.
\label{eq:norm}
\ee
%~~~~~~~~~~~~~~~~~~~~~~~~~~~~~~~~~~~~~~~~~~~~~~~~~~~~~~~~~~~~~~~~~~~~~~~~~~~~
We have omitted the Landau molecular field self-energy in the transport equation,
and we consider the pairing self-energy, $\whDelta$, in the weak-coupling limit,
which is a convenient choice for the order parameter. It
is off-diagonal in particle-hole space,
%~~~~~~~~~~~~~~~~~~~~~~~~~~~~~~~~~~~~~~~~~~~~~~~~~~~~~~~~~~~~~~~~~~~~~~~~~~~~
\be
\whDelta =
\left(\begin{array}{cc}
0 & \hDelta \\
\ul{\hDelta} & 0
\end{array} \right)
=
\left(\begin{array}{cc}
0 & i\hat{\vsigma}\hat{\sigma}_y\cdot\vDelta \\
i\hat{\sigma}_y\hat{\vsigma}\cdot\vDelta^* & 0
\end{array} \right)
\label{eq:dmt}
\ee
%~~~~~~~~~~~~~~~~~~~~~~~~~~~~~~~~~~~~~~~~~~~~~~~~~~~~~~~~~~~~~~~~~~~~~~~~~~~~
and parameterized by a spin-triplet order parameter defined by the vector,
$\vDelta(\hat{\vp},\vR)$. In the weak-coupling limit the order parameter is determined
by the off-diagonal pair amplitude, $\vf(\hat{\vp},\vR;\varepsilon_m)$, from the
gap equation,
%~~~~~~~~~~~~~~~~~~~~~~~~~~~~~~~~~~~~~~~~~~~~~~~~~~~~~~~~~~~~~~~~~~~~~~~~~~~~
\be
\vDelta(\hat{\vp},\vR)=T\sum^{|\varepsilon_m| < \varepsilon_c}_{m}
\int \frac{d\Omega_{\hat\vp'}}{4\pi}\,V(\hat{\vp},\hat{\vp}')\,
                           \vf(\hat{\vp}', \vR; \varepsilon_m)
\,,
\label{eq:selfc}
\ee
%~~~~~~~~~~~~~~~~~~~~~~~~~~~~~~~~~~~~~~~~~~~~~~~~~~~~~~~~~~~~~~~~~~~~~~~~~~~~
where $V(\hat{\vp},\hat{\vp}')$ is the interaction in the spin-triplet
pairing channel. For pure p-wave pairing we retain only the attractive $\ell=1$
interaction, $V=3\,V_{\text{\small 1}}\,\hat{\vp}\cdot\hat{\vp}'$.
The cut-off, $\varepsilon_c$, and interaction, $V_1$, are not measurable, but
they are related to the bulk transition temperature by,
%~~~~~~~~~~~~~~~~~~~~~~~~~~~~~~~~~~~~~~~~~~~~~~~~~~~~~~~~~~~~~~~~~~~~~~~~~~~~
\be
\frac{1}{V_1}=\pi T_c \sum^{|\varepsilon_m|<\varepsilon_c}_{m}{1\over|\varepsilon_m|}
\approx\ln\frac{1.13\varepsilon_c}{T_c}
\,,
\ee
%~~~~~~~~~~~~~~~~~~~~~~~~~~~~~~~~~~~~~~~~~~~~~~~~~~~~~~~~~~~~~~~~~~~~~~~~~~~~
which is used to eliminate the cut-off and pairing interaction in favor
of the measured bulk transition temperature, $T_c$.

The order parameter must be determined self-consistently with the solution of
the transport equation for the propagator. This procedure, and the quasiclassical
transport equation, can be simplified by introducing a parametrization
for the propagator that satisfies the normalization condition by construction
and reduces the number of independent components,
%~~~~~~~~~~~~~~~~~~~~~~~~~~~~~~~~~~~~~~~~~~~~~~~~~~~~~~~~~~~~~~~~~~~~~~~~~~~~
\be
\whg = -i\pi \widehat{N}
\left(\begin{array}{cc}
\hat{1} + \ha\haa & 2\ha \\
 -2\haa & - 1 - \haa \ha
\end{array}\right)
\,,
\label{eq:ricpar}
\ee
%~~~~~~~~~~~~~~~~~~~~~~~~~~~~~~~~~~~~~~~~~~~~~~~~~~~~~~~~~~~~~~~~~~~~~~~~~~
where the prefactor is given by
%~~~~~~~~~~~~~~~~~~~~~~~~~~~~~~~~~~~~~~~~~~~~~~~~~~~~~~~~~~~~~~~~~~~~~~~~~~
\be
\widehat{N} =
\left( \begin{array}{cc}
(1-\ha\haa)^{-1} & 0 \\
0 & (1-\haa\ha)^{-1}
\end{array} \right)
\,.
\label{eq:ricnorm}
\ee
%~~~~~~~~~~~~~~~~~~~~~~~~~~~~~~~~~~~~~~~~~~~~~~~~~~~~~~~~~~~~~~~~~~~~~~~~~~
The amplitudes $\ha$ and $\haa$ are $2\times2$ matrices in spin space which
obey matrix Ricatti equations,\cite{nag93,esc99,esc00}
%~~~~~~~~~~~~~~~~~~~~~~~~~~~~~~~~~~~~~~~~~~~~~~~~~~~~~~~~~~~~~~~~~~~~~~~~~~
\be
\begin{array}{l}
\displaystyle
i\vv_f\cdot\grad \ha + 2i\varepsilon_m \ha - \ha\ul{\hDelta}\ha + \hDelta = 0  \\
\displaystyle
i\vv_f\cdot\grad \haa - 2i\varepsilon_m \haa - \haa\hDelta\ha + \ul{\hDelta} = 0
\,.
\end{array}
\label{eq:ric}
\ee
%~~~~~~~~~~~~~~~~~~~~~~~~~~~~~~~~~~~~~~~~~~~~~~~~~~~~~~~~~~~~~~~~~~~~~~~~~~~~
We refer to $\ha$ and $\haa$ as the Ricatti amplitudes.
The two Ricatti amplitudes are related to the particle- and hole-like projections
of the off-diagonal propagators,
%~~~~~~~~~~~~~~~~~~~~~~~~~~~~~~~~~~~~~~~~~~~~~~~~~~~~~~~~~~~~~~~~~~~~~~~~~~~~
\be
\ha=-(i\pi-\hg)^{-1}\hf\quad,\quad\haa=(i\pi+\ul{\hg})^{-1} \ul{\hf}
\,,
\ee
%~~~~~~~~~~~~~~~~~~~~~~~~~~~~~~~~~~~~~~~~~~~~~~~~~~~~~~~~~~~~~~~~~~~~~~~~~~~~
where the projection operators for the particle- ($\widehat{P}_+$) and hole-like
($\widehat{P}_-$) sectors are given by
%~~~~~~~~~~~~~~~~~~~~~~~~~~~~~~~~~~~~~~~~~~~~~~~~~~~~~~~~~~~~~~~~~~~~~~~~~~~~
\be
\widehat{P}_+ = {1\over 2}\left( 1 + {\whg\over -i\pi}\right)
\quad,\quad
\widehat{P}_- = {1\over 2}\left( 1 - {\whg\over -i\pi}\right)
\,.
\ee
%~~~~~~~~~~~~~~~~~~~~~~~~~~~~~~~~~~~~~~~~~~~~~~~~~~~~~~~~~~~~~~~~~~~~~~~~~~~~
For the case of spin-triplet pairing in zero field these amplitudes
can be parameterized as,
%~~~~~~~~~~~~~~~~~~~~~~~~~~~~~~~~~~~~~~~~~~~~~~~~~~~~~~~~~~~~~~~~~~~~~~~~~~~~
\be
\ha = (i\hat{\vsigma}\cdot\hat{\sigma}_y)\cdot\va
\quad\mbox{and}\quad
\haa = (i\hat{\sigma}_y\hat{\vsigma})\cdot\ul{\va}
\,.
\ee
%~~~~~~~~~~~~~~~~~~~~~~~~~~~~~~~~~~~~~~~~~~~~~~~~~~~~~~~~~~~~~~~~~~~~~~~~~~~~
The Ricatti amplitudes are also related to each other by a symmetry that follows
from symmetry relations for the propagators in Eqs. (\ref{eq:gsm}),
%~~~~~~~~~~~~~~~~~~~~~~~~~~~~~~~~~~~~~~~~~~~~~~~~~~~~~~~~~~~~~~~~~~~~~~~~~~~~
\be
\ha(\hat{\vp}, \vR; \varepsilon_m)^* = \haa(-\hat{\vp}, \vR; \varepsilon_m)
\,.
\label{eq:asm}
\ee
%~~~~~~~~~~~~~~~~~~~~~~~~~~~~~~~~~~~~~~~~~~~~~~~~~~~~~~~~~~~~~~~~~~~~~~~~~~~~

The Ricatti equations are easily integrated numerically,
are numerically stable
and provide a more efficient approach to solving the quasiclassical transport equations
than the ``explosion method''.\cite{thu84}
Equations (\ref{eq:ric}) are solved by integration along classical
trajectories - forward for $\ha$ and backward for $\haa$ - starting from an
initial value. The Ricatti  equations must be supplemented by boundary
conditions at the two interfaces.
We do not have a bulk region in the \He\ film, so generally we start from
an arbitrary initial value at the free surface and compute along a classical
trajectory with multiple reflections until the Ricatti amplitude at the surface converges.
The integration procedure is described in more detail in App. \ref{app:ovch}.

The boundary conditions for the Ricatti amplitudes at the two interfaces are obtained
from boundary conditions for the quasiclassical propagators.
Specular reflection at the free surface requires matching of the
propagators at the free surface for two trajectories, $\hat{\vp}$ and $\ul{\hat{\vp}}$, which
are related by $\ul{\hat{\vp}}=\hat{\vp}-2\hat{\bf n}(\hat{\bf n}\cdot\hat{\vp})$,
%~~~~~~~~~~~~~~~~~~~~~~~~~~~~~~~~~~~~~~~~~~~~~~~~~~~~~~~~~~~~~~~~~~~~~~~~~~~~
\be
\whg(\hat{\vp},D;\varepsilon_m)=\whg(\ul{\hat{\vp}},D;\varepsilon_m) \,.
\label{eq:spec}
\ee
%~~~~~~~~~~~~~~~~~~~~~~~~~~~~~~~~~~~~~~~~~~~~~~~~~~~~~~~~~~~~~~~~~~~~~~~~~~~~
Then the Ricatti amplitudes are also matched at the surface in the same way,
%~~~~~~~~~~~~~~~~~~~~~~~~~~~~~~~~~~~~~~~~~~~~~~~~~~~~~~~~~~~~~~~~~~~~~~~~~~~~
\ber
\ha(\hat{\vp},D;\varepsilon_m)=\ha(\ul{\hat{\vp}},D;\varepsilon_m )
\,,
\\
\haa(\hat{\vp},D;\varepsilon_m)=\haa(\ul{\hat{\vp}},D;\varepsilon_m )
\,.
\label{eq:aspec}
\eer
%~~~~~~~~~~~~~~~~~~~~~~~~~~~~~~~~~~~~~~~~~~~~~~~~~~~~~~~~~~~~~~~~~~~~~~~~~~~~

The boundary condition for the quasiclassical propagator at an atomically rough surface
is more complicated. A physical model for an atomically rough surface is
provided by a ``thin dirty layer'' (TDL) model for surface roughness obtained by coating
a specular surface with a layer (of thickness $d$) of randomly distributed impurities
characterized by a mean free path of $l_{\text{imp}}$.\cite{cul84} In the TDL model the ratio,
$\rho = d/l_{\text{imp}}$, describes the degree of surface roughness. For
$\rho=0$ we recover a specularly reflecting surface, while $\rho \to \infty$ corresponds
to the fully diffuse surface. In the fully diffuse limit we implement Ovchinikov's boundary
condition, which is a special case of the diffuse limit of the TDL boundary condition.
The Ovchinnikov boundary condition requires
self-consistent determination of the Green's function at the diffuse surface,
$\whg(\hat{\vp},0;\varepsilon_m)$. For outgoing trajectories ($\hat{\vp}_z > 0$)
the boundary condition for the Ricatti amplitude is,
%~~~~~~~~~~~~~~~~~~~~~~~~~~~~~~~~~~~~~~~~~~~~~~~~~~~~~~~~~~~~~~~~~~~~~~~~~~~~
\be
\ha(\hat{\vp}, 0) = -(i\pi - \hg_{\text{\tiny TDL}})^{-1} \hf_{\text{\tiny TDL}}
\,,
\label{eq:ovcha0}
\ee
%~~~~~~~~~~~~~~~~~~~~~~~~~~~~~~~~~~~~~~~~~~~~~~~~~~~~~~~~~~~~~~~~~~~~~~~~~~~~
where $\hg_{\text{\tiny TDL}}(\varepsilon_m)$ and $\hf_{\text{\tiny TDL}}(\varepsilon_m)$
are the propagators
deep in the dirty layer, and which are related to the surface propagator by,
%~~~~~~~~~~~~~~~~~~~~~~~~~~~~~~~~~~~~~~~~~~~~~~~~~~~~~~~~~~~~~~~~~~~~~~~~~~~~
\be
\whg_{\text{\tiny TDL}}(\varepsilon_m) =
\int\limits_{{\hat{p}_z>0}\atop{\hat{p}_z<0}} \frac{d\Omega_{\hat{\vp}}}{\pi}
\,|\hat{p}_z|\, \whg(\hat{\vp},0;\varepsilon_m)
\,.
\label{eq:ovg0}
\ee
%~~~~~~~~~~~~~~~~~~~~~~~~~~~~~~~~~~~~~~~~~~~~~~~~~~~~~~~~~~~~~~~~~~~~~~~~~~~~
We give a short derivation of this boundary condition in appendix \ref{app:ovch}.

\subsection*{Order Parameter}

We consider two possible phases in superfluid \He\ films that
have the same, or nearly the same symmetry, as the $A$- and $B$- phases of
bulk supefluid \He\ when one restricts the orbital symmetry group to $SO(2)$.
The order parameter for the B-like phase is of the form,
%~~~~~~~~~~~~~~~~~~~~~~~~~~~~~~~~~~~~~~~~~~~~~~~~~~~~~~~~~~~~~~~~~~~~~~~~~~~~
\be\label{B-phase_OP}
\vDelta_{\text{B}}=(\Delta_\parallel(z)\hat{p}_x, \Delta_\parallel(z) \hat{p}_y,
\Delta_\perp(z)\hat{p}_z)
\,,
\ee
%~~~~~~~~~~~~~~~~~~~~~~~~~~~~~~~~~~~~~~~~~~~~~~~~~~~~~~~~~~~~~~~~~~~~~~~~~~~~
while that for the A-like (`axial') phase is given by,
%~~~~~~~~~~~~~~~~~~~~~~~~~~~~~~~~~~~~~~~~~~~~~~~~~~~~~~~~~~~~~~~~~~~~~~~~~~~~
\be
\vDelta_{\text{A}}=(\,0\,,\,0\,,\Delta_\parallel(z)(\hat{p}_x + i\hat{p}_y))
\,,
\ee
%~~~~~~~~~~~~~~~~~~~~~~~~~~~~~~~~~~~~~~~~~~~~~~~~~~~~~~~~~~~~~~~~~~~~~~~~~~~~
where $\parallel$ and $\perp$ refer to orbital motion, characterized by the direction of the
relative momentum
$\hat{\vp}$, parallel and perpendicular to the surfaces of the film.
The {\sl planar} phase is a special case of the B-phase with $\Delta_\perp=0$.
For the $A$-like
phase we have $\vDelta_{\text{A}} \parallel \vell=\hat{\vz}$ in order to minimize the nuclear
dipolar energy. For the B-phase the order parameter in Eq. (\ref{B-phase_OP}) is multiplied
by a spin-orbit rotation matrix, $\cR(\vn,\vartheta)$, that is fixed by the dipole energy.

\begin{figure}[ht]
\centerline{\psfig{figure=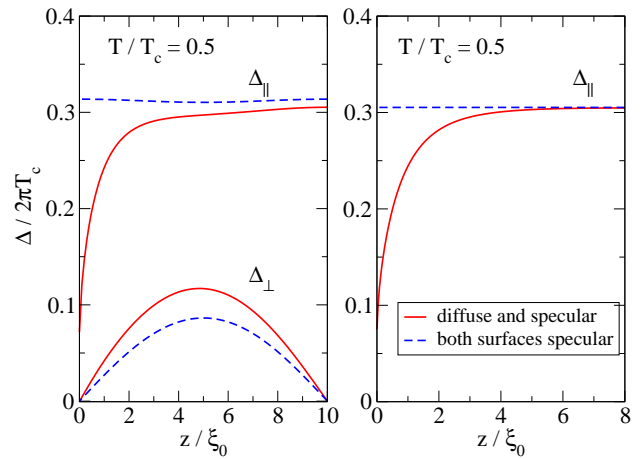,height=6cm}}
\caption{\label{fig:OP}
The order parameters for the $B$-like and the $A$-like phases of \He\ in a thin film.
The coherence length, $\xi_0 = \hbar v_f/2\pi k_B T_c$, is approximately equal
$73\,\text{nm}$ at zero pressure.}
\end{figure}

The spatial profiles of the order parameter components are shown in Fig. \ref{fig:OP}
for both the B- (left panel) and A-like (right panel) phases.
The dashed lines correspond to a film with two specular surfaces, while the solid lines
represent a film with diffuse scattering from a substrate at $z=0$ and specular reflection
from the free surface.

%~~~~~~~~~~~~~~~~~~~~~~~~~~~~~~~~~~~~~~~~~~~~~~~~~~~~~~~~~~~~~~~~~~~~~~~~~~~~~~~~
%~~~~~~~~~~~~~~~~~~~~~~~~~~~~~~~~~~~~~~~~~~~~~~~~~~~~~~~~~~~~~~~~~~~~~~~~~~~~~~~~
The orbital components of the order parameter that are perpendicular to the film
interface, $\Delta_{\perp}$, are suppressed at both interfaces.
This suppression is related to the change of sign of $\Delta_{\perp}\hat{p}_z$
when a quasiparticle is reflected by the surface.
The parallel component, $\Delta_\parallel$, is suppressed
by diffuse scattering at the substrate in both phases. In the $B$-phase
$\Delta_\perp$ is slightly increased for diffuse scattering because some of the
spectral weight that is lost from $\Delta_{\parallel}$ is transferred to $\Delta_{\perp}$.

The orbital structure of the order parameter for the A-phase leads to a simplification
for the boundary condition at the diffuse substrate. If we parameterize the off-diagonal
component of $\whg$ by
%~~~~~~~~~~~~~~~~~~~~~~~~~~~~~~~~~~~~~~~~~~~~~~~~~~~~~~~~~~~~~~~~~~~~~~~~~~~~
\be
\vf=(\,0\,,\,0\,,f_\parallel(\hat{p}_z)(\hat{p}_x+i\hat{p}_y))
\,,
\ee
%~~~~~~~~~~~~~~~~~~~~~~~~~~~~~~~~~~~~~~~~~~~~~~~~~~~~~~~~~~~~~~~~~~~~~~~~~~~~
then the angular integration in Eq. (\ref{eq:ovg0}) gives $\hf_{\text{\tiny TDL}}=0$.
Thus, we have an explicit value for the Ricatti amplitude, $\ha$, at
that surface,
%~~~~~~~~~~~~~~~~~~~~~~~~~~~~~~~~~~~~~~~~~~~~~~~~~~~~~~~~~~~~~~~~~~~~~~~~~~~~
\be
\ha(\hat{\vp},0;\varepsilon_m)= 0\,,\qquad \hat{p}_z >0
\,.
\label{eq:ovchaA}
\ee
%~~~~~~~~~~~~~~~~~~~~~~~~~~~~~~~~~~~~~~~~~~~~~~~~~~~~~~~~~~~~~~~~~~~~~~~~~~~~
The quasiclassical Green's function at $z=0$ is then
%~~~~~~~~~~~~~~~~~~~~~~~~~~~~~~~~~~~~~~~~~~~~~~~~~~~~~~~~~~~~~~~~~~~~~~~~~~~~
\be
\whg(\hat{\vp}, 0, \varepsilon_m) = -i\pi
\left( \begin{array}{cc}
\hat{1}  & 0 \\
-2\haa & - \hat{1}
\end{array}
\right)
\,,
\qquad \hat{p}_z >0
\,.
\ee
%~~~~~~~~~~~~~~~~~~~~~~~~~~~~~~~~~~~~~~~~~~~~~~~~~~~~~~~~~~~~~~~~~~~~~~~~~~~~
The boundary value in Eq. (\ref{eq:ovchaA}) speeds up numerical integration
since the calculation of $\ha(\hat{p}_z>0)$ and
$\haa(\hat{p}_z<0)$ are now initial value problems; we start at $z=0$ with
$\ha(\hat{p}_z>0,0;\varepsilon_m)=0$, or  $\haa(\hat{p}_z<0,0;\varepsilon_m)=0$
and integrate the Ricatti equations directly to obtain
$\ha(\hat{p}_z>0,z;\varepsilon_m)$ and $\haa(\hat{p}_z<0,z;\varepsilon_m)$.

\section{\label{sec:phases} phase diagram}

\begin{figure}[ht]
\centerline{\psfig{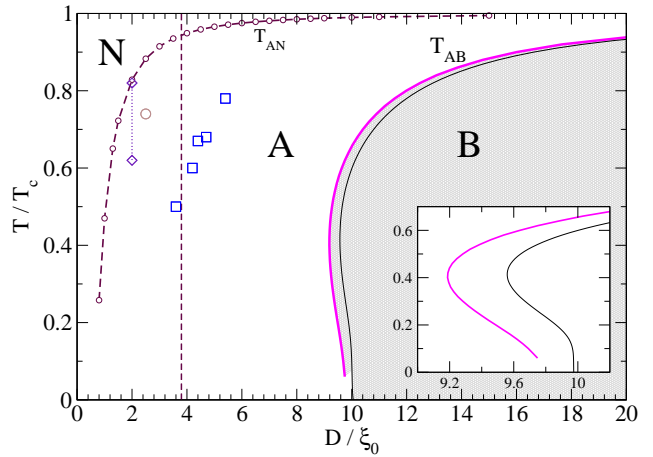}}
\caption{\label{fig:phsdgr} Phase diagram for superfluid \He\ films.
The thick solid line represents the AB transition for a film in contact
with a rough substrate and a specular free surface.
The thin solid line is the AB phase boundary for a film with two specular surfaces.
The inset shows an enlarged portion of the AB phase boundary,
where the second-order transition is re-entrant, A$\rightarrow$B$\rightarrow$A,
as a function of temperature.
The dashed line shows the normal (N) to A-phase boundary, with suppression of
the superfluid transition, $T^{\text{\tiny film}}_c<T_c$, resulting from diffuse scattering by the
substrate. The individual points correspond to observed anomalies in measurements
that may indicate a phase transition in the superfluid film: a) thin dashed line - anomaly
in $\rho_s(T,D)$,\cite{xu90} b) open squares - mode anomaly in third sound,\cite{sch98}
c) open circle - flow anomaly,\cite{ste94} d) open diamonds - flow anomaly.\cite{dav88}}
\end{figure}

At zero pressure bulk \He\ is
in the superfluid B-phase for temperatures below $T_c = 0.93$ mK.
When we confine the superfluid to a slab between two
surfaces, or form a film on a substrate, we observe changes in the superfluid as we
decrease the film thickness, $D$.
The phase diagram of \He\ in superfluid films, as far as it is known, is shown in
Fig. \ref{fig:phsdgr}. Several phase transition lines calculated theoretically are shown,
as well as points indicating possible phase transitions based on anomalies in
several experiments.

If one starts from the bulk superfluid B-phase and then reduces the film thickness, $D$,
at constant temperature we expect to cross at least two phase boundaries.
As $D$ is reduced the perpendicular component, $\Delta_\perp(z)$, is suppressed, and at
a critical film thickness, $D_{\text{\tiny AB}}(T)$, $\Delta_\perp(z)$
vanishes. This signifies a transition to the {\sl planar} phase with an
order parameter of the form $\vDelta_{\text{P}}=\Delta_\parallel(z)
(\hat{p}_x\,,\hat{p}_y\,,\,0\,)$. The component $\Delta_\perp(z)$ vanishes continuously,
so this transition is second order. In the weak-coupling limit the planar and axial phases
are degenerate. However, in bulk \He\ strong-coupling corrections lower the free energy
of the axial phase relative to the planar phase. Thus, if strong-coupling effects also
stabilize the axial
phase relative to the planar phase in a thin film then the second-order transition from
the B-phase to the planar phase is pre-empted by a transition from the
B-phase to the axial A-phase. Measurements of the heat capacity jump in bulk \He\ indicate that
strong-coupling corrections are small at zero pressure, thus, the AB transition in
the film is likely to be very weakly first order. With the
exception of possible fine structure of the phase diagram close to the second-order
transition line, $D_{\text{\tiny AB}}(T)$,
and properties such as the latent heat of transition, calculations of the thermodynamic
properties of thin films based on the weak-coupling approximation are expected to be accurate.

The perpendicular component of the order parameter is suppressed to zero even for a specular
wall, so the AB transition occurs even in a film bounded by two specular surfaces.
For a film on a rough substrate, the suppression of $\Delta_\parallel$ by diffuse
scattering leads to a small enhancement of $\Delta_\perp(z)$.
As a result the B to A transition requires slightly thinner films for a rough substrate.
This result, although the detailed shape of the phase boundary is slightly different, agrees with
the calculations reported by Nagato \et\cite{nag00} based on
a different theoretical model for the surface roughness.
However, NMR measurements\cite{kaw98} on thin slabs of superfluid \He\
show that the AB transition occurs at larger values
of film thickness than predicted by the weak-coupling theory.
This may indicate that the first-order AB phase boundary needs to
be calculated with leading order strong-coupling corrections included
in the theory, even at zero pressure.

%~~~~~~~~~~~~~~~~~~~~~~~~~~~~~~~~~~~~~~~~~~~~~~~~~~~~~~~~~~~~~~~~~~~~~~~~~~~~~~~~
%~~~~~~~~~~~~~~~~~~~~~~~~~~~~~~~~~~~~~~~~~~~~~~~~~~~~~~~~~~~~~~~~~~~~~~~~~~~~~~~~
An interesting feature of the calculated weak-coupling AB phase boundary below
$T/T_c \approx 0.4$ is shown in the inset of Fig. \ref{fig:phsdgr}. For films
in contact with either a specular or rough substrate the second-order phase
boundary is re-entrant as a function of temperature for a narrow range of film
thicknesses.  For example, for a film on a rough surface with $D\simeq
9.4\xi_0$, upon decreasing the temperature below $T_c^{\text{\tiny film}}$ the
A to B transition occurs at $T_{\text{\tiny AB}}\approx 0.55 T_c$. As the
temperature drops further a re-entrant B to A transition occurs at
$T_{\text{\tiny BA}}\simeq 0.23T_c$. Whether or not this re-entrance will
survive strong-coupling corrections is not known.
The re-entrance may also
signal that a translationally invariant A or B phase is unstable to formation
of an inhomogeneous phase with lower free energy.
In any event the fine structure of the phase diagram at low
temperatures near $D\simeq 9.5\xi_0$, and the possibility of new phases
stabilized by strong-coupling corrections, or which spontaneously break
translation symmetry in the plane of the film, is outside the scope of this
paper.

For films that are thinner than $9.5-10\,\xi_0$ the planar or axial A-phase is the
stable phase relative to the B-like phase. For the purpose of calculating the
thermodynamic properties we assume that strong-coupling corrections stabilize the A-phase
relative to planar phase in the film, however, this is really an open question.
Strong-coupling corrections to the
free energy for phases with strong spatial variations, as occurs in thin films, have
not been calculated, so the relative stability of the planar and A-phase in thin films is
unknown, either theoretically or experimentally.

If the substrate were an ideal specularly reflecting
surface then the superfluid A-phase would persist for film thicknesses
approaching a few monolayers, or until the Fermi liquid properties and pairing interaction
were modified by finite size effects. But, the pair-breaking effect of scattering off
a rough substrate suppresses transition temperature into the A-phase
and at a film width $D_{\text{\tiny AN}}(T)$, which is substantially smaller than
$D_{\text{\tiny AB}}(T)$ the superfluid A-phase is destroyed.
The calculated transition temperature for diffuse scattering is shown
in Fig. \ref{fig:phsdgr}. This phase boundary was calculated by identifying
the temperature and film thickness where the order parameter vanishes.
We also obtained the superfluid transition from the calculated
free energy by a least square fit of the known Ginzburg-Landau form for the
free energy, $a(T - T^{\text{\tiny film}}_c)^2$.

Calculations of the transition temperature in thin slabs of
\He\ were carried out by Kj{\"a}ldman \et\cite{kja87} using a linearized gap equation
and deGennes' formulation of the kernel in terms of the classical limit for the normal-state
current-current correlation function.\cite{amb75}
Our calculations agree well with the results for a slab if we take into account that the width of
the thin film of \Hea\ is equivalent to a \Hea\ slab of twice the width of the film.

It should be noted that the phases considered here, even for thin \He\ films, assume
thicknesses, $D\gg\text{\AA}$. We do not consider the two-dimensional limit of
one or two atomic layers of \He\ atoms on the surface of a substrate. The properties of
2D superfluid \He, if it exists, are expected to be influenced by the reduced dimensionality.
Ising-like, as well as Kosterlitz-Thouless-type transitions are predicted for
2D superfluid \Hea \cite{ste79a, kaw99}.

One additional note:
observing the equilibrium phase boundaries may be complicated by metastability.
Even though the planar phase, i.e. the B-phase with $\Delta_{\perp}=0$,
and the axial A-phase are degenerate
in weak-coupling they are unrelated by symmetry, and therefore, separated by an energy barrier.
Thus, once established, the axial A-phase will be metastable with respect to the B-phase.
The calculation of the barrier and the corresponding metastability lines in the phase diagram
would provide an important result, but are outside the scope of this article.

\section{\label{sec:dos} Density of states}

Pair-breaking by surface scattering leads to quasiparticle states
below the gap. These excitations play an important role in the
thermodynamic and transport properties of thin films of superfluid
\He. The sub-gap excitations are surface Andreev bound states. The
mechanism leading to their formation is closely related to the
formation of bound states in the core of a vortex.\cite{rai96}
Andreev bound state formation occurs when the order parameter
changes sign or phase along a quasiparticle trajectory. In the
case of surface scattering the incident and reflected trajectories
generally correspond to very different order parameters; this is
typically the case for an unconventional order parameter which
breaks rotational symmetry.

For example, consider the B-phase in contact with a specular surface. For an
incident trajectory normal to the interface, $\hat{\vp}\parallel\hat{\vz}$ the
B-phase order parameter changes sign upon reflection, i.e.
$\vDelta(\ul{\hat\vp})=-\vDelta(\hat\vp)$ for $\hat\vp\rightarrow \ul{\hat\vp}=-\hat\vp$. This
sign change leads to multiple Andreev reflection that generates a surface
bound state at the Fermi level, i.e. $\varepsilon=0$.
For trajectories away from normal incidence the components of the
B-phase order parameter corresponding to orbital motion in the
plane are present and do not change sign upon reflection. As a
result the surface Andreev bound state disperses as a function of
the incident and reflected angles relative to the interface normal.

%~~~~~~~~~~~~~~~~~~~~~~~~~~~~~~~~~~~~~~~~~~~~~~~~~~~~~~~~~~~~~~~~~~~~~~~~~~~~~~~~
\begin{figure}[ht]
\centerline{\psfig{figure=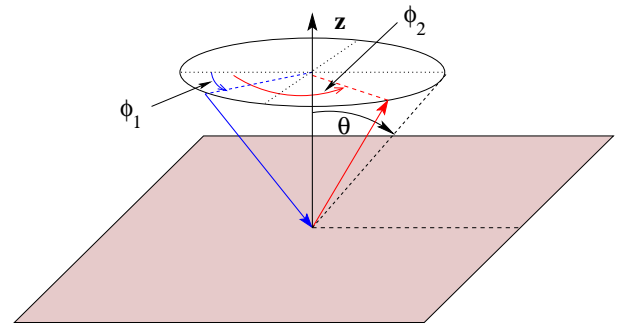,width=8cm}}
\caption{\label{fig:skew_scattering}
Skew scattering by the substrate.}
\end{figure}
%~~~~~~~~~~~~~~~~~~~~~~~~~~~~~~~~~~~~~~~~~~~~~~~~~~~~~~~~~~~~~~~~~~~~~~~~~~~~~~~~
The bound state energy dispersion can be calculated approximately by neglecting
the suppression of the order parameter at the surface, and assuming that surface
scattering occurs on a cone defined by the angle $\theta$ from the $xy$-plane,
i.e. $(\theta, \phi_1) \to (\pi - \theta, \phi_2)$, $\phi = \phi_2 - \phi_1$
as shown in Fig. \ref{fig:skew_scattering}.
We then find bound-state poles in the retarded propagator, for either the B- or A-phase,
given by
%~~~~~~~~~~~~~~~~~~~~~~~~~~~~~~~~~~~~~~~~~~~~~~~~~~~~~~~~~~~~~~~~~~~~~~~~~~~~~~~~
\be
\varepsilon_{\text{b}}=\pm\Delta_\parallel\,\sin\theta\,\cos{\phi\over 2}
\,.
\label{eq:bs}
\ee
%~~~~~~~~~~~~~~~~~~~~~~~~~~~~~~~~~~~~~~~~~~~~~~~~~~~~~~~~~~~~~~~~~~~~~~~~~~~~~~~~

The density of states (DOS) can be calculated once the order parameter and Landau
molecular fields have been determined self-consistently. The most detailed information
is contained in the angle-resolved local density of states, which is obtained from the
diagonal component of the retarded quasiclassical propagator,
%~~~~~~~~~~~~~~~~~~~~~~~~~~~~~~~~~~~~~~~~~~~~~~~~~~~~~~~~~~~~~~~~~~~~~~~~~~~~~~~~
\be
\text{N}(\hat{\vp},\vR;\varepsilon)=-\frac{1}{\pi}\Im g^{R}(\hat{\vp},\vR;\varepsilon)
\ee
%~~~~~~~~~~~~~~~~~~~~~~~~~~~~~~~~~~~~~~~~~~~~~~~~~~~~~~~~~~~~~~~~~~~~~~~~~~~~~~~~
where $g^{R}(\hat{\vp}, \vR; \varepsilon)$ is found by solving the quasiclassical
transport equation for real energies; i.e. $i\varepsilon_m\to\varepsilon + i0^+$
and the known order parameter and molecular fields.
The local density of states for the B-phase near a wall shows quasiparticle
states which develop below the bulk gap, and are bound to the surface, i.e. their spectral
weight vanishes a few coherence lengths away from the surface.

%~~~~~~~~~~~~~~~~~~~~
\begin{figure}[ht]
\centerline{\psfig{figure=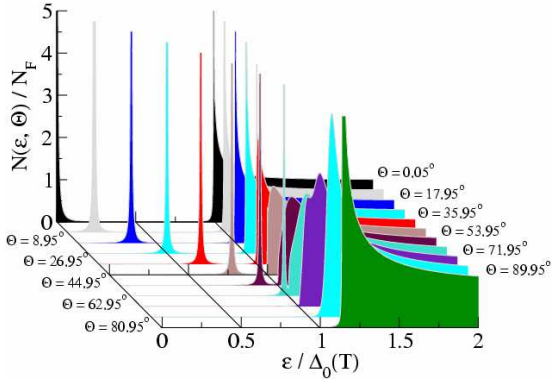,height=5cm}}
\caption{\label{fig:dosBspec}
The angle-resolved local DOS for the \Heb\ near a specular surface. The spectrum is
calculated for $T=0.5T_c$. For clarity we have broadened the Andreev bound states with a width
parameter, $\eta = 10^{-3}\Delta_0$.
}
\end{figure}
%~~~~~~~~~~~~~~~~~~~~

For example, the angle-resolved spectrum of superfluid \Heb\ near a specular surface,
calculated numerically for a self-consistently determined order parameter,
is shown in Fig. \ref{fig:dosBspec}. For the specular reflection the position of the
positive energy surface bound state depends on the angle of the incident
trajectory, $\theta$, approximately as
$\varepsilon_{\text{\tiny b}} =\Delta_\parallel \,\sin\theta$.
For normal incidence the bound state is at zero energy
and disperses towards, eventually merging into, the continuum edge as the
incident trajectory approaches grazing incidence. There is also a weak
dispersion in the continuum edge reflecting the enhancement of $\Delta_{\perp}$ by
surface scattering.

%~~~~~~~~~~~~~~~~~~~~
\begin{figure}[ht]
\centerline{\psfig{figure=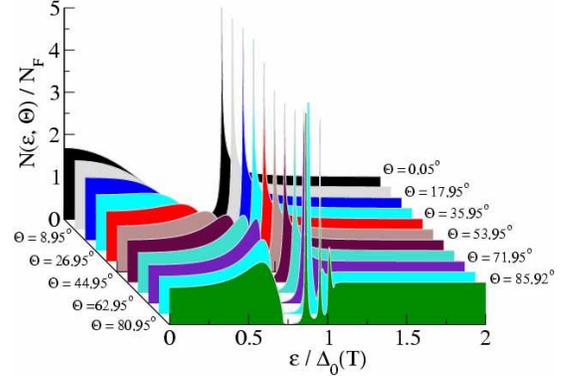,height=5cm}}
\caption{\label{fig:dosBdiffuse}
The angle-resolved local DOS for the \Heb\ near a diffuse surface. The spectrum is
calculated for $T=0.5T_c$. For clarity we have broadened the Andreev bound states
near grazing incidence with a width parameter of $\eta = 10^{-3}\Delta_0$.
}
\end{figure}
%~~~~~~~~~~~~~~~~~~~~

At an atomically rough surface diffuse scattering couples an incident trajectory to
all outgoing trajectories. This leads to mixing of states with different energies
and thus to a band of sub-gap states for a given incident trajectory as shown
in Fig. \ref{fig:dosBdiffuse}. The suppression of $\Delta_\parallel$ for diffuse
scattering also leads to the formation of additional subgap states bound by
multiple Andreev reflection within the ``pair potential'' provided by the
suppressed order parameter, $\Delta_\parallel(z)$. These states appear only near
grazing incidence and are weakly bound with energies just below the continuum edge.

\begin{figure}[ht]
\centerline{\psfig{figure=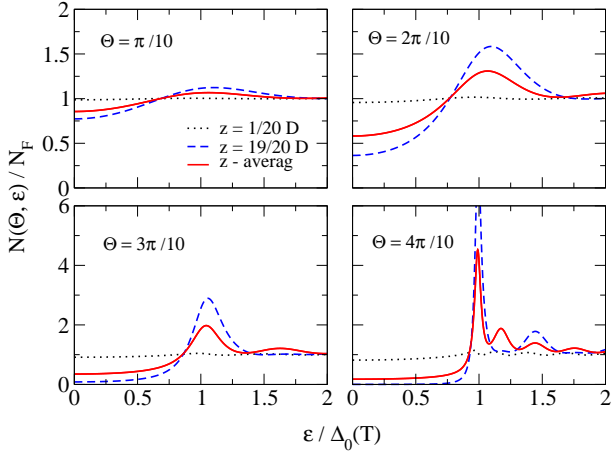,height=6cm}}
\caption{\label{fig:dosAang}
The density of states in the A-phase for a film of thickness $D = 4\, \xi_0$ at
temperature $T = 0.01 T_c$. The excitation energy is scaled in units of $\Delta_0(T)$,
the bulk value of $\Delta_\parallel$ at temperature $T$.
}
\end{figure}

Sub-gap states do not appear in \Hea\ at a specular wall since there is no
change in phase of the order parameter for specular reflection when
$\vell\parallel\hat\vz$. Thus, all quasiparticle states belong to the continuum.
This situation changes dramatically for a rough surface.
Now there are scattering processes connecting an incident trajectory with a reflected
trajectory that is at a skew angle, $\phi\ne\pi$, in the $xy$-plane
(see Fig. \ref{fig:skew_scattering}). For \Hea\ the order parameter for a trajectory,
$\hat\vp=\cos\theta\hat\vz + \sin\theta(\cos\phi\hat\vx+\sin\phi\hat\vy)$, is
$\vDelta_{\text{\tiny A}}(\hat\vp)=\hat\vz\,\Delta_\parallel\sin\theta\,e^{i\phi}$.
The change in phase of the order parameter upon skew scattering leads to strong pair-breaking
and the formation of sub-gap states. For the diffuse scattering the coupling of skew
trajectories with all possible azimuthal angles generates a band of states which fill
the sub-gap spectrum as shown in Fig. \ref{fig:dosAang} for several incident trajectories.

%~~~~~~~~~~~~~~~~~~~~~~~~~~~~~~~~~~~~~~

%~~~~~~~~~~~~~~~~~~~~~~~~~~~~~~~~~~~~~~

The self-consistent spectrum calcuated numerically and shown in Fig. \ref{fig:dosAang}
is well described by the spectrum obtained by calculating the retarded
propagator for a constant order parameter, $\Delta_\parallel$, everywhere in the film.
The spectrum is then
determined entirely by the changes in the order parameter induced by diffuse scattering.
The transport equation can be solved analytically with Ovchinnikov's boundary condition
for the Riccati amplitudes. For the diagonal part of the quasiclassical propagator we obtain,

%~~~~~~~~~~~~~~~~~~~~~~~~~~~~~~~~~~~~~~~~~~~~~~~~~~~~~~~~~~~~~~~~~~~~~~~~~~~~~~~~
\begin{widetext}
\be
g(\theta, z; \varepsilon_m)=-i \pi \left(1 +
\frac{1}{1 +
\frac{\displaystyle \varepsilon_m}{\displaystyle \Delta_\parallel^2 \sin^2\theta}
\left(\varepsilon_m + \omega \tanh [2\omega D/v_f \cos\theta]\right)}
\left( \frac{\cosh \, [2\omega (z-D)/v_f \cos \theta]}
{\cosh \, [2\omega D/v_f \cos \theta]} - 1 \right) \right) \,,
\ee
%~~~~~~~~~~~~~~~~~~~~~~~~~~~~~~~~~~~~~~~~~~~~~~~~~~~~~~~~~~~~~~~~~~~~~~~~~~~~~~~~
\end{widetext}
where
$\omega^2=\Delta(\theta)^2+\varepsilon_m^2$ with
$\Delta(\theta)=\Delta_\parallel\,\sin\theta$.
We calculate the retarded propagator by analytic continuation to the real energy axis,
$i\varepsilon_m\rightarrow\varepsilon+i0^+$. The result for the local density of states
is then,
\begin{widetext}
%~~~~~~~~~~~~~~~~~~~~~~~~~~~~~~~~~~~~~~~~~~~~~~~~~~~~~~~~~~~~~~~~~~~~~~~~~~~~~~~~
\ber\label{eq:dosAanalytic}
\text{N}(\theta, z;\varepsilon)/\text{N}_f = 1 +
\Theta(\varepsilon^2 - \Delta(\theta)^2)
\frac{1}{1 -
\frac{\displaystyle \varepsilon^2}{\displaystyle \Delta(\theta)^2}
\left(1 + \tan^2 [2\omega D/v_f \cos\theta]\right)}
\left( \frac{\cos \, [2\omega (z-D)/v_f \cos \theta]}
{\cos \, [2\omega D/v_f \cos \theta]} - 1 \right) \nonumber \\
+ \Theta(\Delta(\theta)^2-\varepsilon^2)
\frac{1}{1 -
\frac{\displaystyle \varepsilon^2}{\displaystyle \Delta(\theta)^2}
\left(1 - \tanh^2 [2\omega D/v_f \cos\theta]\right)}
\left( \frac{\cosh \, [2\omega (z-D)/v_f \cos \theta]}
{\cosh \, [2\omega D/v_f \cos \theta]} - 1 \right) \,,
\eer
%~~~~~~~~~~~~~~~~~~~~~~~~~~~~~~~~~~~~~~~~~~~~~~~~~~~~~~~~~~~~~~~~~~~~~~~~~~~~~~~~
\end{widetext}
%~~~~~~~~~~~~~~~~~~~~~~~~~~~~~~~~~~~~~~~~~~~~~~~~~~~~~~~~~~~~~~~~~~~~~~~~~~~~~~~~
where $\omega$ is now
%~~~~~~~~~~~~~~~~~~~~~~~~~~~~~~~~~~~~~~~~~~~~~~~~~~~~~~~~~~~~~~~~~~~~~~~~~~~~~~~~
\be
\omega = \sqrt{|\varepsilon^2 - \Delta(\theta)^2|} \,.
\ee
%~~~~~~~~~~~~~~~~~~~~~~~~~~~~~~~~~~~~~~~~~~~~~~~~~~~~~~~~~~~~~~~~~~~~~~~~~~~~~~~~
Equation (\ref{eq:dosAanalytic}) shows both the band of the sub-gap states generated
by diffuse scattering and the Tomasch oscillations for energies above the continuum. The
positions of the maxima are determined by the condition for constructive interference
of particle- and hole-like excitations with energies above the gap,
$\vert\Delta_\parallel\sin\theta\vert$, reflecting from the specular surface,
%~~~~~~~~~~~~~~~~~~~~~~~~~~~~~~~~~~~~~~~~~~~~~~~~~~~~~~~~~~~~~~~~~~~~~~~~~~~~~~~~
\be
\frac{2\sqrt{\varepsilon^2 - \Delta_\parallel^2 \sin^2\theta}}{v_f \cos \theta} D
= n\,\pi\quad,\quad n=0,1,2\dots
\,.
\ee
%~~~~~~~~~~~~~~~~~~~~~~~~~~~~~~~~~~~~~~~~~~~~~~~~~~~~~~~~~~~~~~~~~~~~~~~~~~~~~~~~
The spectral weight of the Tomasch oscillations also depends on distance from the surface.
Some peaks are suppressed at special positions in the film due to spatial oscillations
of the particle-hole interference amplitudes. This suppression of spectral weight is
most visible for angles close to $\theta = \pi/2$ near the the free surface.
For example, the density of states for $\theta = 0.4\pi$ at $z=19/20 D$
shows that every second peak is suppressed.

\begin{figure}[ht]
\centerline{\psfig{figure=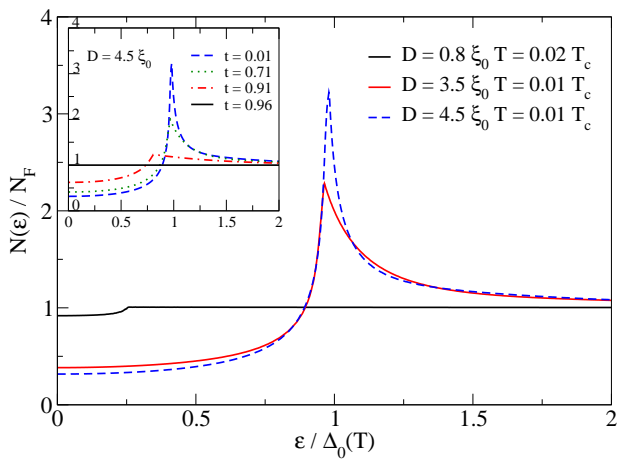,height=6cm}}
\caption{\label{fig:dosAav}
The total DOS averaged over film for several film thicknesses for $T\ll T_c^{\text{\tiny film}}$. The inset
shows the gapless excitations filling the gap for $t=T/T_c^{\text{\tiny film}}\to 1$.
}
\end{figure}

The sub-gap states are bound to the surface on the length scale set by
coherence length, $\xi_0$, and decay exponentially into the bulk. However,
the situation is different for thin films; the bound states extend over the
entire width of the film. Figure \ref{fig:dosAav} shows the total DOS averaged
over the film,
$\text{N}(\varepsilon)=\int\,dz/D\,\int d\Omega_{\hat\vp}/4\pi\,\text{N}(\hat{\vp},z;\varepsilon)$.
The spectrum is gapless over the entire energy range, $\varepsilon < \Delta_0$,
and is finite at $\varepsilon = 0$. The inset to Fig. \ref{fig:dosAav} shows the
evolution of the total DOS as a function of temperature. The gapless states fill the spectrum
$\varepsilon < \Delta_0$ completely as $T \to T_c^{\text{\tiny film}}$.

At low temperatures, $T \to 0$, the DOS is insensitive to temperature, and the value
of the DOS at $\varepsilon=0$, $0 < \text{N}(0) < \text{N}_f$, persists above $\varepsilon=0$.
If we decrease the film thickness the sub-gap states fill the gap
and the transition to the normal state will occur when this process is complete.
As shown in Fig. \ref{fig:dosAav} the density of states is almost
equal to that for the normal state over the whole energy range for $D=0.8\xi_0$.
The A to normal transition
occurs for a slightly smaller film thickness.

As a consequence of the gapless spectrum the thermodynamic properties of the films of \Hea\ will
be very different from those of bulk \He. For example, the low-temperature behavior of the
specific heat vanishes exponentially in bulk \Heb\ for $T\ll\Delta_{\text{\tiny B}}$,
$C_{\text{\tiny B}}^{\text{\tiny bulk}}(T)\to(T_c/T)^{3/2}\,\exp(-\Delta_0/T)$, while for
bulk \Hea\ the nodal excitations have zero energy at isolated points on the Fermi surface.
The bulk density of states vanishes at the Fermi level as
$\text{N}_{\text{\tiny A}}^{\text{\tiny bulk}}(\varepsilon\to0) \sim \varepsilon^2$,
and the specific heat exhibits a power law,
$C_{\text{\tiny A}}^{\text{\tiny bulk}}\sim(T/T_c)^{3}$.

The specific heat of films of superfluid \Hea\ is expected to have a
different power law behavior at low temperatures.
The density of states is finite and nearly constant in the low-energy range above
the Fermi level. As a result the
specific heat will have the linear temperature dependence as $T\to 0$, just as for normal
\He, except that the Sommerfeld coefficient is reduced in the superfluid film by the
ratio, $\text{N}(0)/\text{N}_f$.

%~~~~~~~~~~~~~~~~~~~~~~~~~~~~~~~~~~~~~~~~~~~~~~~~~~~~~~~~~~~~~~~~~~~~~~~~~~~~~~~
%~~~~~~~~~~~~~~~~~~~~~~~~~~~~~~~~~~~~~~~~~~~~~~~~~~~~~~~~~~~~~~~~~~~~~~~~~~~~~~~

\section{\label{sec:fe} Thermodynamic properties}

To compute the thermodynamic properties of superfluid \He\ films we need a free
energy functional formulated in terms of the quasiclassical propagator and self-energies.
Such a functional has been derived starting from the general Luttinger-Ward functional,
formulated in terms of the full Green's function and self-energy, by eliminating the
high-energy, short-wavelength intermediate states, and thus computing
only corrections to the ground-state energy to leading order in the small expansion
parameters of Fermi liquid theory. The conceptual formulation of this problem is discussed
in detail by Rainer and Serene.\cite{rai76} The formulation of a quasiclassical free-energy
functional for inhomogeneous equilibrium states is similar, but their are additional technical
approaches to incorporating inhomogeneities of the order parameter.\cite{ser83,thu84,buc95a}
Our approach is similar to that of Ref. \onlinecite{ser83} and is outlined
in App. \ref{app:fe}.

We start from the quasiclassical free energy in the weak-coupling limit expressed in terms
of the quasiclassical propagator, $\whg(\hat{\vp},\vR;\varepsilon_m)$ and
order parameter, $\whDelta(\hat{\vp},\vR)$, derived in Eq. (\ref{eq:a2-FE}),
%~~~~~~~~~~~~~~~~~~~~~~~~~~~~~~~~~~~~~~~~~~~~~~~~~~~~~~~~~~~~~~~~~~~~~~~~~~~~~~~~
\be
\Del\Omega[\whg,\whDelta]=
-\onefourth\text{Sp}'(\whDelta\,\whg)
+ \onehalf\int_0^1 d\lambda\,\text{Sp}'(\whDelta\whg_\lambda)
\,.
\label{eq:fe}
\ee
%~~~~~~~~~~~~~~~~~~~~~~~~~~~~~~~~~~~~~~~~~~~~~~~~~~~~~~~~~~~~~~~~~~~~~~~~~~~~~~~~
The symbol, $\mbox{Sp}'$, denotes the sum over relevant variables:
the volume of the \He\ film, position on the Fermi surface, Matsubara energies and a
trace over spin and particle-hole degrees of freedom,
%~~~~~~~~~~~~~~~~~~~~~~~~~~~~~~~~~~~~~~~~~~~~~~~~~~~~~~~~~~~~~~~~~~~~~~~~~~~~~~~~
\be
\text{Sp}'(\ldots) =\text{N}_f\int d^3 R\,\int\frac{d\Omega_{\hat\vp}}{4\pi}\,
T\sum_{m}\,\text{Tr}_4(\ldots)
\,.
\ee
%~~~~~~~~~~~~~~~~~~~~~~~~~~~~~~~~~~~~~~~~~~~~~~~~~~~~~~~~~~~~~~~~~~~~~~~~~~~~~~~~
There is an additional integration over the variable coupling parameter $\lambda$ in Eq. (\ref{eq:fe})
involving an auxillary propagator, $\whg_\lambda$, which is the
solution of the quasiclassical transport equation in Eq. (\ref{eq:transport_equation}),
but with the self-energy scaled by $\lambda$: $\whDelta\to\whDelta_\lambda=\lambda\whDelta$.
The transport equation for $\whg_\lambda$ is not solved self-consistently, but with
a single integration for each value of $\lambda$. Thus, $\whg_\lambda$ is a function of the
exact order parameter in the film. This procedure and the application
of boundary conditions for computing the auxillary propagator and the quasiclassical
free energy functional are also explained in App. \ref{app:fe}.

%~~~~~~~~~~~~~~~~~~~~~~
\begin{figure}[ht]
\centerline{\psfig{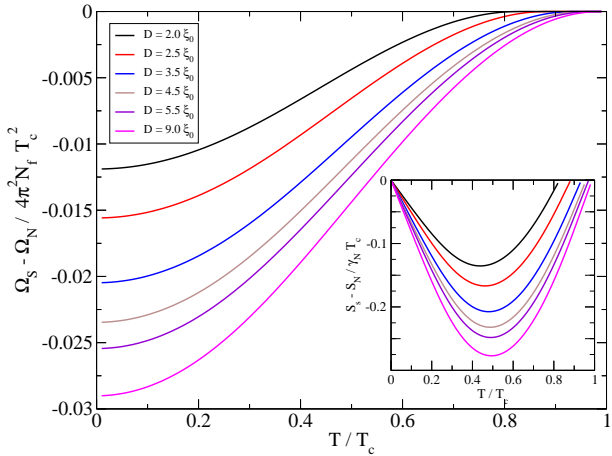}}
\caption{\label{fig:FE}
Superfluid correction to the thermodynamic potential vs. reduced temperature for several films
of superfluid \Hea\ on an rough substrate. The inset shows the reduction in
the entropy of the superfluid film.
}
\end{figure}
%~~~~~~~~~~~~~~~~~~~~~~

Equation (\ref{eq:fe}), when evaluated with the self-consistently determined propagator and order
parameter in the film, gives the difference of the thermodynamic potential,
$\Del\Omega=\Omega_{\text{\tiny S}}-\Omega_{\text{\tiny N}}$, from which the change
in entropy and specific heat can be calculated,
%~~~~~~~~~~~~~~~~~~~~~~~~~~~~~~~~~~~~~~~~~~~~~~~~~~~~~~~~~~~~~~~~~~~~~~~~~~~~~~~
\be
\Del S(T) = -\frac{\partial\Del\Omega}{\partial T}
\quad,\quad
\Del C(T) = -T\frac{\partial^2\Del\Omega}{\partial T^2}
\,.
\ee
%~~~~~~~~~~~~~~~~~~~~~~~~~~~~~~~~~~~~~~~~~~~~~~~~~~~~~~~~~~~~~~~~~~~~~~~~~~~~~~~

The normal-state free energy for \He\ of volume $V$ is given by
$\Omega_{\text{\tiny N}}(T)=E_{\text{\tiny N}}-
V(\tinyonehalf\gamma_{\text{\tiny N}}T^2)$,
where $E_{\text{\tiny N}}$ is the ground-state energy for normal \He\ and
$\gamma_{\text{\tiny N}}=\frac{2\pi^2}{3}\text{N}_f\,k_{\text{\tiny B}}^2$ is the
normal-state Sommerfeld coefficient.

The reduction of the free energy
below the normal-state value represents the gain in energy due to the
formation of a condensate of pairs in the film.
The free energy of \Hea\ films in the limit of diffuse scattering by the substrate
is shown in Fig. \ref{fig:FE} for several film thicknesses. The reduction in the free energy
is given by the Ginzburg-Landau form, $\Del\Omega\propto-(1-T/T_c^{\text{\tiny film}})^2$,
for temperatures just below the superfluid transition temperature, $T_c^{\text{\tiny film}}$,
of the film. At low temperatures the gapless excitations dominate the
thermodynamics. The density of states at the Fermi energy is non-zero and approximately
constant at low energies. As a result the low-temperature limit for the
free-energy of the superfluid state decreases as
$\Omega_{\text{\tiny S}}-E_{\text{\tiny S}}=-V\,(\tinyonehalf\gamma_{\text{\tiny S}}T^2)$,
where $\gamma_{\text{\tiny S}}<\gamma_{\text{\tiny N}}$ is the Sommerfeld coefficient
for the low-energy excitations of the superfluid film, and $E_{\text{\tiny S}}$ is the
$T=0$ condensation energy.

From the numerical results shown in Fig. \ref{fig:FE} we
can calculate temperature dependence of the entropy and specific heat of the \Hea\ film.
The results for the entropy are shown in the inset of Fig. \ref{fig:FE}. The linear temperature
dependence of the entropy resulting from the gapless excitations is clearly visible.

\begin{figure}[t]
\centerline{\psfig{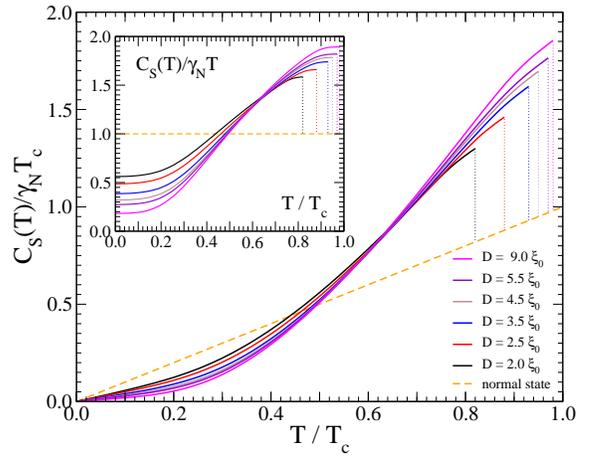}}
\caption{\label{fig:C}
The specific heat of a superfluid \Hea\ film as a function of reduced temperature
for several film thicknesses. The inset shows the ratio,
$C_{\text{\tiny S}}(T)/C_{\text{\tiny N}}(T)$.
}
\end{figure}

Numerical calculations of the specific heat are shown in Fig. \ref{fig:C}. These results
show the decrease of the heat capacity jump at $T_c^{\text{\tiny film}}$ with decreasing film thickness, as
well as the linear temperature dependence of the specific heat resulting from the gapless
excitations (see inset of Fig. \ref{fig:FE}).
This behavior for \Hea\ films is in sharp contrast to the low temperature heat capacity
of bulk \Hea, which varies as $C_{\text{\tiny S}}\sim T^3$.

%~~~~~~~~~~~~~~~~~~~~~~~~~
%~~~~~~~~~~~~~~~~~~~~~~~~~

\begin{figure}[ht]
\centerline{\psfig{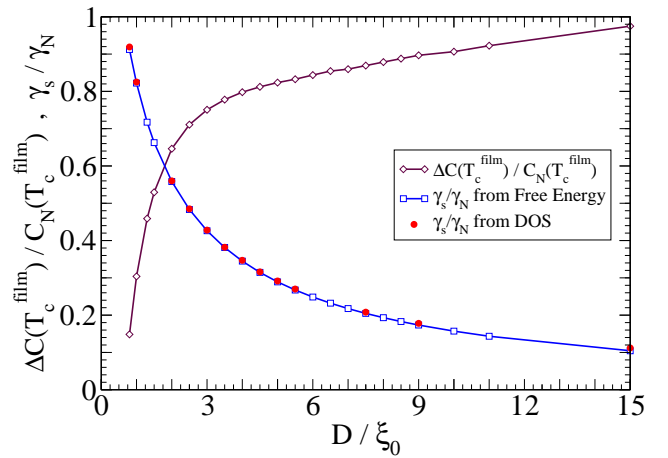}}
\caption{\label{fig:cjump}
Specific heat jump, $\Del C(T_c^{\text{\tiny film}})/\gamma_{\text{\tiny N}}T_c^{\text{\tiny film}}$
at $T_c^{\text{\tiny film}}$ and the ratio of the low-temperature Sommerfeld coefficient,
$\gamma_{\text{\tiny S}}/\gamma_{\text{\tiny N}}$, for the superfluid film as a function of film
thickness, $D$. For comparison, $\Del C(T_c)/\gamma_{\text{\tiny N}}T_c=1.19$ for bulk \Hea\ in the
weak-coupling limit.
}
\end{figure}

Results for the heat capacity jump, $\Del C(T_c^{\text{\tiny film}})/\gamma_N T_c^{\text{\tiny film}}$,
and the Sommerfeld coefficient, $\gamma_{\text{\tiny S}}$,
are summarized in Fig. \ref{fig:cjump} as a function of the film thickness, $D$.
The Sommerfeld coefficient, $\gamma_{\text{\tiny S}}$, was calculated by two independent
methods. We calculated $\gamma_{\text{\tiny S}}$ directly by numerically differentiating
the temperature dependence of the superfluid free energy. We can also relate the
Sommerfeld coefficient directly to the density of states at the Fermi energy, $\text{N}(0)$.
Thus, $\gamma_{\text{\tiny S}}/\gamma_{\text{\tiny N}}=\text{N}(0)/\text{N}_f$. The first
calculation is carried out entirely in the Matsubara formalism, while the calculation of the
DOS at the Fermi level is obtained by the solving for the retarded quasiclassical propagator
on the real energy axis. Both results agree and are shown in Fig. \ref{fig:cjump}, and give
us confidence in our numerical calculations for the propagators, free energy, entropy
and heat capacity.

%~~~~~~~~~~~~~~~~~~~~~~~~~~~~~~

%~~~~~~~~~~~~~~~~~~~~~~~~~~~~~~

\section{\label{sec:cnclsn} Conclusion}

We have calculated the thermodynamic properties of thin films of superfluid \He\
in the weak-coupling limit, expected to be applicable to films at zero pressure.
We calculated the phase diagram for the superfluid film, including the AB transition,
the suppression of the superfluid transition temperature, suppression of the
order parameter, the quasiparticle density of states and thermodynamic potential.
Our analysis, based on the quasiclassical method, shows a spectrum for
superfluid films with gapless excitations formed by the combination of reflection
by a rough substrate and Andreev scattering induced by changes in the order parameter along
classical trajectories of quasiparticles.
The gapless excitation spectrum depends on the film thickness and dominates the
low temperature thermodynamic potential, entropy and specific heat.

\begin{acknowledgments}
We thank Matthias Eschrig and Tomas L{\"o}fwander for helpful discusssions,
and acknowledge support from NSF through grant DMR-9972087.
\end{acknowledgments}

\appendix

\section{\label{app:ovch} Diffuse Scattering}

The boundary condition for the quasiclassical propagator at a rough surface is obtained
by solving the transport equation,
%~~~~~~~~~~~~~~~~~~~~~~~~~~~~~~~~~~~~~~~~~~~~~~~~~~~~~~~~~~~~~~~~~~~~~~~~~~~~~~~~~~~~~~~
\ber
\big[i\varepsilon_m \widehat{\tau}_3-\whsig_{\text{\tiny imp}}(\hat\vp,\vR;\varepsilon_m)
- \whDelta(\hat{\vp},\vR),\whg(\hat{\vp},\vR;\varepsilon_m)\big]
\nonumber \\
+ i\vv_f\cdot\grad\whg(\hat{\vp},\vR;\varepsilon_m) = 0
\,,
\eer
%~~~~~~~~~~~~~~~~~~~~~~~~~~~~~~~~~~~~~~~~~~~~~~~~~~~~~~~~~~~~~~~~~~~~~~~~~~~~~~~~~~~~~~~
in the dirty layer and matching this solution to the quasiclassical propagator in the
superfluid. In the limit of strong disorder within the impurity layer,
$|\mbox{\footnotesize $\mathfrak{S}$}_{\text{\tiny imp}}|\gg|\varepsilon_m|,|\Del|$.
Thus, deep in the impurity layer,
%~~~~~~~~~~~~~~~~~~~~~~~~~~~~~~~~~~~~~~~~~~~~~~~~~~~~~~~~~~~~~~~~~~~~~~~~~~~~~~~~~~~~~~~
\be
\left[\whsig_{\text{\tiny imp}}(\hat{\vp},\vR;\varepsilon_m),
      \whg(\hat{\vp},\vR;\varepsilon_m)\right] = 0
\,,
\label{eq:a1-0}
\ee
%~~~~~~~~~~~~~~~~~~~~~~~~~~~~~~~~~~~~~~~~~~~~~~~~~~~~~~~~~~~~~~~~~~~~~~~~~~~~~~~~~~~~~~~
with the impurity self-energy evaluated in the Born approximation,
%~~~~~~~~~~~~~~~~~~~~~~~~~~~~~~~~~~~~~~~~~~~~~~~~~~~~~~~~~~~~~~~~~~~~~~~~~~~~~~~~~~~~~~~
\be
\whsig_{\text{\tiny imp}} =
{1\over 2\pi}\int{d\Omega_{\hat{\vp}'}\over 4\pi}\tau^{-1}(\hat{\vp},\hat{\vp}')
\whg(\hat{\vp}',\vR;\varepsilon_m)
\,,
\label{eq:a1-2}
\ee
%~~~~~~~~~~~~~~~~~~~~~~~~~~~~~~~~~~~~~~~~~~~~~~~~~~~~~~~~~~~~~~~~~~~~~~~~~~~~~~~~~~~~~~~
where $\tau^{-1}(\hat{\vp},\hat{\vp}')$ is the rate for quasiparticles
to scatter from $\hat{\vp} \to \hat{\vp}'$ on the Fermi surface.
These equations are solved by an isotropic propagator, $\whg_{\text{\tiny TDL}}(\varepsilon_m)$,
that is normalized to $\whg_{\text{\tiny TDL}}^2=-\pi^2\hat{1}$. This propagator
is {\sl not} the normal-state propagator for the isolated normal metal because the proximity
coupling to the superfluid layer produces a `rotation' of
$\whg_{\text{\tiny N}}(\varepsilon_m)\to\whg_{\text{\tiny TDL}}(\varepsilon_m)$ in particle-hole
space. To fix this `rotation' we include the leading corrections to Eq. (\ref{eq:a1-0})
due to spatial variations of the propagator in both the TDL and the superfluid film and
match the solutions at the interface. In the TDL the transport equation is
%~~~~~~~~~~~~~~~~~~~~~~~~~~~~~~~~~~~~~~~~~~~~~~~~~~~~~~~~~~~~~~~~~~~~~~~~~~~~~~~~~~~~~~~
\be
\hspace*{-1em}i\vv_f\cdot\grad\whg(\hat{\vp},\vR;\varepsilon_m) =
[\whsig_{\text{\tiny imp}}(\hat{\vp},\vR;\varepsilon_m),
\whg(\hat{\vp},\vR;\varepsilon_m)]
\,.
\label{eq:a1-1}
\ee
%~~~~~~~~~~~~~~~~~~~~~~~~~~~~~~~~~~~~~~~~~~~~~~~~~~~~~~~~~~~~~~~~~~~~~~~~~~~~~~~~~~~~~~~
Equations (\ref{eq:a1-2}-\ref{eq:a1-1}) are solved by expanding the propagator in a basis
of Nambu matrices.
For superfluid \He\ films in zero field the basis is limited to matrices in $2\times 2$
particle-hole space, with the spin degrees of freedom fixed. Thus,
three linearly independent matrices, $\{\whg_1,\whg_2,\whg_3\}$, are required (the
identity matrix drops out of Eq. (\ref{eq:a1-1}). These matrices satisfy
the algebraic relations of the Pauli matrices,
%~~~~~~~~~~~~~~~~~~~~~~~~~~~~~~~~~~~~~~~~~~~~~~~~~~~~~~~~~~~~~~~~~~~~~~~~~~~~~~~~~~~~~~~
\be
[\whg_i,\whg_j]_{+} = -2 \pi^2 \, \delta_{ij}
\,,\quad
\left[ \whg_i,\whg_j\right]_{-}=-2\pi\,\varepsilon_{ijk}\,\whg_k
\,.
\label{eq:a1-comm}
\ee
%~~~~~~~~~~~~~~~~~~~~~~~~~~~~~~~~~~~~~~~~~~~~~~~~~~~~~~~~~~~~~~~~~~~~~~~~~~~~~~~~~~~~~~~
We choose $\whg_3=\whg_{\text{\tiny TDL}}$ and express the propagator in the TDL as,
%~~~~~~~~~~~~~~~~~~~~~~~~~~~~~~~~~~~~~~~~~~~~~~~~~~~~~~~~~~~~~~~~~~~~~~~~~~~~~~~~~~~~~~~
\ber
\whg(\hat{\vp}, \vR; \varepsilon_m) =
  B_+(\hat{\vp}, \vR) \, \whg_+(\varepsilon_m)
+ B_-(\hat{\vp}, \vR) \, \whg_-(\varepsilon_m)\nonumber\\
+ B_3(\hat{\vp}, \vR) \, \whg_3(\varepsilon_m)\,.
\eer
%~~~~~~~~~~~~~~~~~~~~~~~~~~~~~~~~~~~~~~~~~~~~~~~~~~~~~~~~~~~~~~~~~~~~~~~~~~~~~~~~~~~~~~~
where $\whg_\pm=(\whg_1\pm i\whg_2)/\sqrt{2}$. The linear differential equations
for $\{ B_3(\hat{\vp},\vR),B_+(\hat{\vp}, \vR),B_-(\hat{\vp},\vR)\}$ are easily solved
with Ovchinnikov's model of forward scattering,
$\tau^{-1}(\hat{\vp},\hat{\vp}')= 4\tau^{-1}\hat{p}_z\hat{p}'_z$ for
$\hat{p}_z\hat{p}'_z>0$, otherwise $\tau^{-1}(\hat{\vp},\hat{\vp}')= 0$.
Thus, quasiparticles enter the dirty layer, scatter forward towards the specular wall,
and after reflection diffuse out of the TDL.
The limit: $d\to 0$, $v_f\tau\to 0$, $v_f\tau/d\to 0$ corresponds to diffuse scattering
by the impurity layer.

The propagator in the TDL is matched to the propagator, $\whg(\hat\vp,0;\varepsilon_m)$, in
the superfluid at the interface to the TDL. We use the same basis to express
%~~~~~~~~~~~~~~~~~~~~~~~~~~~~~~~~~~~~~~~~~~~~~~~~~~~~~~~~~~~~~~~~~~~~~~~~~~~~~~~~~~~~~~~
\ber
\whg(\hat{\vp},0;\varepsilon_m)=\whg_{\text{\tiny TDL}}(\varepsilon_m)
&+&
C_+(\hat{\vp},0)\whg_+(\varepsilon_m)
\nonumber \\
&+&
C_-(\hat{\vp}, 0)\whg_-(\varepsilon_m)
\,.
\label{eq:ovchg}
\eer
%~~~~~~~~~~~~~~~~~~~~~~~~~~~~~~~~~~~~~~~~~~~~~~~~~~~~~~~~~~~~~~~~~~~~~~~~~~~~~~~~~~~~~~~
The coefficients of this expansion satisfy the following
relations obtained by Ovchinnikov,\cite{ovc69}
%~~~~~~~~~~~~~~~~~~~~~~~~~~~~~~~~~~~~~~~~~~~~~~~~~~~~~~~~~~~~~~~~~~~~~~~~~~~~~~~~~~~~~~~
\ber
C_+(\hat{\vp}, 0) = 0\,,\quad\hat{p}_z &<&0\,,\quad\text{and} \\
& & \int_{\hat{p}_z>0} \frac{d\Omega_{\hat{p}}}{\pi}|\hat{p}_z| C_+(\hat{\vp}, 0) = 0 \,,\nonumber \\
C_-(\hat{\vp}, 0) = 0\,,\quad\hat{p}_z &>& 0\,,\quad\text{and}\label{eq:ovchC}         \nonumber\\
& & \int_{\hat{p}_z<0} \frac{d\Omega_{\hat{p}}}{\pi}|\hat{p}_z| C_-(\hat{\vp}, 0) = 0 \,.\nonumber
\eer
%~~~~~~~~~~~~~~~~~~~~~~~~~~~~~~~~~~~~~~~~~~~~~~~~~~~~~~~~~~~~~~~~~~~~~~~~~~~~~~~~~~~~~~~
The propagator deep in the dirty layer
is also related to the physical propagator at the boundary,
%~~~~~~~~~~~~~~~~~~~~~~~~~~~~~~~~~~~~~~~~~~~~~~~~~~~~~~~~~~~~~~~~~~~~~~~~~~~~~~~~~~~~~~~
\be
\whg_{\text{\tiny TDL}}(\varepsilon_m) =
\int\limits_{\hat{p}_z>0 \atop \hat{p}_z<0} \frac{d\Omega_{\hat{p}}}{\pi}
\,|\hat{p}_z|\, \whg(\hat{\vp}, 0, \varepsilon_m) \,.
\label{eq:ovchg0}
\ee
%~~~~~~~~~~~~~~~~~~~~~~~~~~~~~~~~~~~~~~~~~~~~~~~~~~~~~~~~~~~~~~~~~~~~~~~~~~~~~~~~~~~~~~~
Boundary conditions, \ref{eq:ovchg}-\ref{eq:ovchg0}, can be written in a compact form
using the commutation relations (\ref{eq:a1-comm}),
%~~~~~~~~~~~~~~~~~~~~~~~~~~~~~~~~~~~~~~~~~~~~~~~~~~~~~~~~~~~~~~~~~~~~~~~~~~~~~~~~~~~~~~~
\be\begin{array}{lll}
\whg(\hat{\vp},0;\varepsilon_m) & -\,\,\whg_{\text{\tiny TDL}}(\varepsilon_m) =  & \\
                                & \displaystyle{\frac{\mbox{sgn}(\hat{p}_z)}{2\pi i}
 [\whg_{\text{\tiny TDL}}(\varepsilon_m)\,,\,\whg(\hat{\vp},0;\varepsilon_m)]}   &
\,.\nonumber
\label{eq:ovchcomm}
\end{array}
\ee
%~~~~~~~~~~~~~~~~~~~~~~~~~~~~~~~~~~~~~~~~~~~~~~~~~~~~~~~~~~~~~~~~~~~~~~~~~~~~~~~~~~~~~~~
This condition is solved self-consistently with Eq. (\ref{eq:ovchg0}) for $\whg_{\text{\tiny TDL}}$
and $\whg(\hat\vp,0;\varepsilon_m)$.

The boundary condition for $\whg(\hat\vp,0;\varepsilon_m)$ can be cast into a more compact
form using the Ricatti representation for the propagator, $\whg$.
For an outgoing trajectory, $\hat{p}_z > 0$,
Eq. (\ref{eq:ovchcomm}) is solved by
%~~~~~~~~~~~~~~~~~~~~~~~~~~~~~~~~~~~~~~~~~~~~~~~~~~~~~~~~~~~~~~~~~~~~~~~~~~~~~~~~~~~~~~~
\be
\ha(\hat{\vp},0)=-(i\pi-\hg_{\text{\tiny TDL}})^{-1} \hf_{\text{\tiny TDL}}
\,.
\label{eq:ovcha}
\ee
%~~~~~~~~~~~~~~~~~~~~~~~~~~~~~~~~~~~~~~~~~~~~~~~~~~~~~~~~~~~~~~~~~~~~~~~~~~~~~~~~~~~~~~~
Thus, integration along an outgoing trajectory should start
with the value of $\ha$ given by the value deep in the thin dirty layer.
The second Ricatti amplitude, $\haa$, is known at the TDL substrate, since we integrate $\haa$
along a trajectory with $\hat{p}_z > 0$ in backward direction. Thus, equations
(\ref{eq:ovchg0}) and (\ref{eq:ovcha}), together with
$\haa(\hat{\vp},0;\varepsilon_m)$ and the Ricatti parametrization (\ref{eq:ricpar}-\ref{eq:ricnorm})
are iterated until they converge to a value for $\ha(\hat{\vp},0)$.

\begin{figure}[ht]
\centerline{\psfig{figure=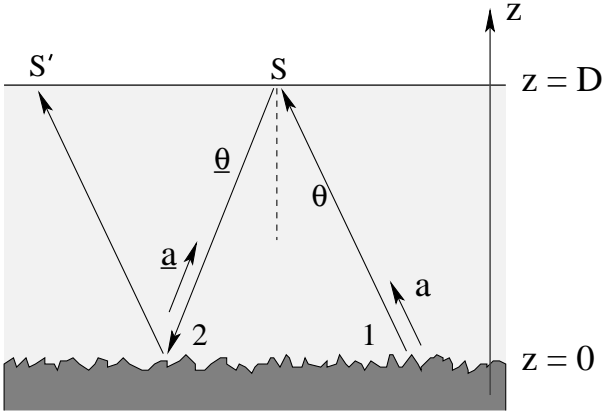,width=8cm}}
\caption{\label{fig:integr} Integration along classical
trajectories.}
\end{figure}

We use the fourth-order Runge-Kutta method to numerically integrate the Ricatti equations
along a classical trajectory for the Ricatti amplitudes, $\ha(\vp,z;\varepsilon_m)$ and
$\haa(\vp,z;\varepsilon_m)$. Azimuthal symmetry for scattering in the plane of the film
allows us to consider trajectories defined by $\theta$ and $\phi=0$ (Fig. \ref{fig:integr}).
%~~~~~~~~~~~~~~~~~~~~~~~~~~~~~~~~~~~~~~~~~~~~~~~~~~~~~~~~~~~~~~~~~~~~~~~~~~~~~~~
%~~~~~~~~~~~~~~~~~~~~~~~~~~~~~~~~~~~~~~~~~~~~~~~~~~~~~~~~~~~~~~~~~~~~~~~~~~~~~~~
The integration procedure is slightly different for the A- and B-phases. In case of the A-phase
we know the Riccati amplitudes at the interface with the substrate (Eq. \ref{eq:ovchaA}):
$\ha(\theta,0;\varepsilon_m)=\haa(\underline{\theta},0;\varepsilon_m)=0$. For any
trajectory we start at point $1$ and integrate forward along trajectory $1-S-2$,
with specular reflection at $S$, to obtain the amplitude $\ha$. To calculate $\haa$ we
integrate along the same trajectory in the reverse direction starting at point $2$.

For the B-phase we do not know the initial values of the Ricatti amplitudes anywhere. In this
case we start with an initial guess for the amplitude $\ha$, e.g. at the point $S$. Using
Eq. \ref{eq:asm} and inversion in the azimuthal plane we find a starting value
for the amplitude $\haa$. We then integrate from $S$ to $2$ to obtain
$\ha(\underline{\theta},0;\varepsilon_m)$, and from $S'$ to $2$ to obtain
$\haa(\theta,0;\varepsilon_m)$. We implement the diffuse boundary conditions at the point
$2$ to obtain $\ha(\theta, 0;\varepsilon_m)$, and then integrate from $2$ to $S'$. This gives
us an updated initial value for the amplitude $\ha$ (and by symmetry for $\haa$).
The integration procedure is repeated until convergence is reached.

\section{\label{app:fe} Free energy functional}

We start with the Luttinger-Ward functional for the full Nambu
Green's function, $\whG$, and self-energy, $\whS$,
%~~~~~~~~~~~~~~~~~~~~~~~~~~~~~~~~~~~~~~~~~~~~~~~~~~~~~~~~~~~~~~~~~~~~~~~~~~~~~~~~~~~~~~~~~~~~
\be
\Omega[\whG, \whS]=-{1\over 2}\text{Sp}
\left\{\whS\,\whG+\ln(-\whG^{-1}_0+\whS)\right\}+\Phi[\whG]
\,,\label{eq:a2-lw}
\ee
%~~~~~~~~~~~~~~~~~~~~~~~~~~~~~~~~~~~~~~~~~~~~~~~~~~~~~~~~~~~~~~~~~~~~~~~~~~~~~~~~~~~~~~~~~~~~
where
%~~~~~~~~~~~~~~~~~~~~~~~~~~~~~~~~~~~~~~~~~~~~~~~~~~~~~~~~~~~~~~~~~~~~~~~~~~~~~~~~~~~~~~~~~~~~
\be
\text{Sp}\left\{\ldots\right\} = T \sum_{\varepsilon_m}\,
\int d^3R\,\int\frac{d^3 p}{(2\pi)^3}\,
\mbox{Tr}_4\left\{\ldots\right\}
\,,
\ee
%~~~~~~~~~~~~~~~~~~~~~~~~~~~~~~~~~~~~~~~~~~~~~~~~~~~~~~~~~~~~~~~~~~~~~~~~~~~~~~~~~~~~~~~~~~~~
$\whS(\vp,\vR;\varepsilon_m)$ is the self-energy and
$\whG^{-1}_0(\vp,\vR;\varepsilon_m)=i\varepsilon_m\widehat{\tau}_3-
 \xi_{\text{\tiny 0}}(\vp)\widehat{1}$
is the inverse Green's function for non-interacting reference system of bare \He.
The stationarity condition with respect to the Green's function,
%~~~~~~~~~~~~~~~~~~~~~~~~~~~~~~~~~~~~~~~~~~~~~~~~~~~~~~~~~~~~~~~~~~~~~~~~~~~~~~~~~~~~~~~~~~~~
\be
\frac{\delta \Omega}{\delta
\whG^{\text{\tiny tr}}}=0
\rightsquigarrow
\whS=\whS_{\text{\tiny skel}}[\whG]=
2\frac{\delta\Phi[\whG]}{\delta\whG^{\text{\tiny tr}}}
\,,
\ee
%~~~~~~~~~~~~~~~~~~~~~~~~~~~~~~~~~~~~~~~~~~~~~~~~~~~~~~~~~~~~~~~~~~~~~~~~~~~~~~~~~~~~~~~~~~~~
relates the functional $\Phi[\whG]$ to the self-energy via the skeleton expansion for the
self-energy, while the stationarity condition with respect to the self-energy,
%~~~~~~~~~~~~~~~~~~~~~~~~~~~~~~~~~~~~~~~~~~~~~~~~~~~~~~~~~~~~~~~~~~~~~~~~~~~~~~~~~~~~~~~~~~~~
\be
\frac{\delta \Omega}{\delta \whS^{\text{\tiny tr}}} = 0
\rightsquigarrow
\whG^{-1} = \whG^{-1}_0 - \whS
\,,
\ee
%~~~~~~~~~~~~~~~~~~~~~~~~~~~~~~~~~~~~~~~~~~~~~~~~~~~~~~~~~~~~~~~~~~~~~~~~~~~~~~~~~~~~~~~~~~~~
gives Dyson's equation for the Nambu propagator. These equations provide a starting point for
deriving the Fermi-liquid theory for superfluid \He. In particular, the leading order expansion of
Dyson's equation in the small parameters of Fermi liquid theory can be transformed
into Eilenberger's transport equation (Eq. \ref{eq:transport_equation}) for the quasiclassical propagator.

In order to derive a free energy functional of the quasiclassical
propagator, $\whg$, and quasiclassical self-energy,
%~~~~~~~~~~~~~~~~~~~~~~~~~~~~~~~~~~~~~~~~~~~~~~~~~~~~~~~~~~~~~~~~~~~~~~~~~~~~~~~~~~~~~~~~~~~~
\be
\whsig(\hat{\vp},\vR;\varepsilon_m)\equiv
a\left(\whS(p_f\hat{\vp},\vR;\varepsilon_m)-\whS_{\text{\tiny N}}\right)
\widehat{\tau}_3
\,,
\ee
%~~~~~~~~~~~~~~~~~~~~~~~~~~~~~~~~~~~~~~~~~~~~~~~~~~~~~~~~~~~~~~~~~~~~~~~~~~~~~~~~~~~~~~~~~~~~
we remove the normal-state stationary point ($\whG_{\text{\tiny N}},\whS_{\text{\tiny N}}$)
from the Luttinger-Ward functional
(Eq. \ref{eq:a2-lw}) by defining $\Del\whS=\whS-\whS_{\text{\tiny N}}$,
$\Del\whG = \whG - \whG_{\text{\tiny N}}$ and introducing the subtracted functional,
%~~~~~~~~~~~~~~~~~~~~~~~~~~~~~~~~~~~~~~~~~~~~~~~~~~~~~~~~~~~~~~~~~~~~~~~~~~~~~~~~~~~~~~~~~~~~
\ber
\lefteqn{\Del\Omega[\whG,\Del\whS]
= \Omega[\whG,\whS] - \Omega[\whG_{\text{\tiny N}},\whS_{\text{\tiny N}}] = }
\nonumber\\
&&- \onehalf\,\text{Sp}\,
\left\{\Del\whS \,\whG + \ln(-\whG^{-1}_{\text{\tiny N}}+\Del\whS)
                       - \ln(-\whG^{-1}_{\text{\tiny N}})
\right\}
\nonumber\\
&&+ \Del\Phi[\whG]
\,,
\label{eq:a2-lwN}
\eer
%~~~~~~~~~~~~~~~~~~~~~~~~~~~~~~~~~~~~~~~~~~~~~~~~~~~~~~~~~~~~~~~~~~~~~~~~~~~~~~~~~~~~~~~~~~~~
which has as inputs the normal-state propagator,
$\whG_{\text{\tiny N}} = (\whG_0^{-1} - \whS_{\text{\tiny N}})^{-1}$,
and self-energy, $\whS_{\text{\tiny N}}$, rather than the bare propagator.
The subtracted $\Phi$-functional,
%~~~~~~~~~~~~~~~~~~~~~~~~~~~~~~~~~~~~~~~~~~~~~~~~~~~~~~~~~~~~~~~~~~~~~~~~~~~~~~~~~~~~~~~~~~~~
\be
\Del\Phi[\whG] = \Phi[\whG] - \Phi[\whG_{\text{\tiny N}}] - {1\over 2} \text{Sp}
\left\{ \whS_{\text{\tiny N}} \, (\whG - \whG_{\text{\tiny N}}) \right\}
\,,
\ee
%~~~~~~~~~~~~~~~~~~~~~~~~~~~~~~~~~~~~~~~~~~~~~~~~~~~~~~~~~~~~~~~~~~~~~~~~~~~~~~~~~~~~~~~~~~~~
is confined to the low-energy region of phase space since pairing
corrections to the normal-state propagator contribute
only in the low-energy region, $k_BT_c\ll E_f$. The diagrammatic perturbation expansion
for $\Del\Phi$ can be reorganized as an asymptotic expansion in the small parameters of
Fermi liquid theory\cite{rai76} that is formally an expansion in the number of low-energy
propagator lines.

To convert Eq. (\ref{eq:a2-lwN}) to a functional of the quasiclassical propagator
and self-energy we integrate out the momentum dependence normal to the Fermi surface
over a region of momentum space near the Fermi surface,
$\vert\xi_{\vp}\vert < \varepsilon_c$. The low-energy self-energy is slowly varying function
of $\xi_{\vp}$ and can be evaluated with $\vp=p_f\hat{\vp}$. Thus, the term,
$\Delta\whS\,\whG$, in Eq. (\ref{eq:a2-lwN}) is $\xi_p$-integrated to give
%~~~~~~~~~~~~~~~~~~~~~~~~~~~~~~~~~~~~~~~~~~~~~~~~~~~~~~~~~~~~~~~~~~~~~~~~~~~~~~~~~~~~~~~~~~~~
\ber
\text{Sp} \left\{\Delta\whS\,\whG \right\}\Rightarrow
\text{Sp}'\left\{\whsig\,\whg\right\}\equiv&&
\\
N_f\,T\sum_m\int d^3 R\int\frac{d\Omega_{\hat\vp}}{4\pi}
\mbox{Tr}_4
&\big(\whsig(\hat\vp,\vR;\varepsilon_m)\whg(\hat\vp,\vR;\varepsilon_m)\big)\,.&
\nonumber
\eer
%~~~~~~~~~~~~~~~~~~~~~~~~~~~~~~~~~~~~~~~~~~~~~~~~~~~~~~~~~~~~~~~~~~~~~~~~~~~~~~~~~~~~~~~~~~~~

To integrate the $\ln$-functional we introduce an auxiliary
functional defined by introducing a variable coupling constant for the self-energy
and $\Phi$-functional: $\Del\whS\to\Del\whS_\lambda\equiv\lambda\Del\whS$ and
$\Del\Phi\to\Del\Phi_\lambda\equiv\lambda\Del\Phi$. Thus, the auxiliary functional is
%~~~~~~~~~~~~~~~~~~~~~~~~~~~~~~~~~~~~~~~~~~~~~~~~~~~~~~~~~~~~~~~~~~~~~~~~~~~~~~~~~~~~~~~~~~~~
\ber
\lefteqn{\Del\Omega_\lambda[\whG,\Del\whS_\lambda]= }& &
\nonumber\\
&-&{1\over 2}\text{Sp}
\left\{\Del\whS_\lambda\,\whG+\ln(-\whG^{-1}_{\text{\tiny N}}+\Del\whS_\lambda)
                             -\ln(-\whG^{-1}_{\text{\tiny N}})\right\}
\nonumber\\
& + & \lambda \, \Del\Phi[\whG]
\,.
\label{eq:a2-lwL}
\eer
%~~~~~~~~~~~~~~~~~~~~~~~~~~~~~~~~~~~~~~~~~~~~~~~~~~~~~~~~~~~~~~~~~~~~~~~~~~~~~~~~~~~~~~~~~~~~
The stationarity conditions with respect to $\whG$ and $\Delta\whS_\lambda$
give a new equation for an auxiliary propagator,
$\whG^{-1}_\lambda \equiv \whG^{-1}_{\text{\tiny N}}-\Delta \whS_\lambda$.
The auxiliarly functional can be $\xi_{\vp}$-integrated after first differenting
with respect to the coupling parameter,
then carrying out the $\xi_{\vp}$-integration to obtain,
%~~~~~~~~~~~~~~~~~~~~~~~~~~~~~~~~~~~~~~~~~~~~~~~~~~~~~~~~~~~~~~~~~~~~~~~~~~~~~~~~~~~~~~~~~~~~
\be
\frac{\partial\Del\Omega_\lambda}{\partial\lambda} =
    -\onehalf\text{Sp}'\left\{\whsig\,\whg \right\} +
     \onehalf\text{Sp}'\left\{\whsig\,\whg_\lambda \right\} + \Del\Phi[\whg]
\,,
\label{eq:a2-lwLQ}
\ee
%~~~~~~~~~~~~~~~~~~~~~~~~~~~~~~~~~~~~~~~~~~~~~~~~~~~~~~~~~~~~~~~~~~~~~~~~~~~~~~~~~~~~~~~~~~~~
where
\ber
\whg_{\lambda}(\hat{\vp},\vR;\varepsilon_m)
= \frac{1}{a}\int_{-\varepsilon_c}^{+\varepsilon_c}\,d\xi_{\vp}
    \widehat{\tau}_3\,\whG_{\lambda}(\vp,\vR;\varepsilon_m)
\,,
\eer
%~~~~~~~~~~~~~~~~~~~~~~~~~~~~~~~~~~~~~~~~~~~~~~~~~~~~~~~~~~~~~~~~~~~~~~~~~~~~~~~~~~~~~~~~~~~~
is the quasiclassical auxiliary propagator.
We can integrate Eq. (\ref{eq:a2-lwLQ}) with respect to the coupling constant. Since
$\Del\Omega_{\lambda=0}=0$ and $\Del\Omega_{\lambda=1}=\Del\Omega$
we obtain the desired free energy functional in terms of the quasiclassical
propagator, $\whg$, and self-energy, $\whsig$,
%~~~~~~~~~~~~~~~~~~~~~~~~~~~~~~~~~~~~~~~~~~~~~~~~~~~~~~~~~~~~~~~~~~~~~~~~~~~~~~~~~~~~~~~~~~~~
\be
\Del\Omega[\whg,\whsig] =
\onehalf\int_0^1\,d\lambda\,\text{Sp}'\left\{\whsig\,\left(\whg_\lambda-\whg\right)\right\}+
\Del\Phi[\whg]
\,.
\label{eq:a2-lwQ}
\ee
%~~~~~~~~~~~~~~~~~~~~~~~~~~~~~~~~~~~~~~~~~~~~~~~~~~~~~~~~~~~~~~~~~~~~~~~~~~~~~~~~~~~~~~~~~~~~

The stationarity conditions for the subtracted free-energy functional
reduce to the quasiclassical transport equation and self-energy expansion obtained
from the asymptotic expansion of the $\Phi$ functional,
%~~~~~~~~~~~~~~~~~~~~~~~~~~~~~~~~~~~~~~~~~~~~~~~~~~~~~~~~~~~~~~~~~~~~~~~~~~~~~~~~~~~~~~~~~~~~
\be
[i\varepsilon_m\widehat{\tau}_3-\whsig,\whg] + i\vv_f\cdot\grad\whg = 0
\,,\qquad
\whsig = 2\frac{\delta\Del\Phi[\whg]}{\delta\whg^{\text{\tiny tr}}}
\,.
\ee
%~~~~~~~~~~~~~~~~~~~~~~~~~~~~~~~~~~~~~~~~~~~~~~~~~~~~~~~~~~~~~~~~~~~~~~~~~~~~~~~~~~~~~~~~~~~~
These equations are supplemented by boundary conditions
for the propagator $\whg$ which describe the effects of scattering by
a surface or interface.

The auxiliary propagator, $\whg_{\lambda}$, is a functional of the exact quasiclassical
self-energy, and is obtained by solving the quasiclassical transport equation
with $\whsig\rightarrow\lambda\whsig$,

%~~~~~~~~~~~~~~~~~~~~~~~~~~~~~~~~~~~~~~~~~~~~~~~~~~~~~~~~~~~~~~~~~~~~~~~~~~~~~~~~~~~~~~~~~~~~
\be\label{eq:a2-auxiliary_transport}
\left[i\varepsilon_m\widehat{\tau}_3-\lambda\,\whsig,\whg_\lambda\right]
+i\vv_f\cdot\grad\whg_\lambda = 0
\,.
\ee
%~~~~~~~~~~~~~~~~~~~~~~~~~~~~~~~~~~~~~~~~~~~~~~~~~~~~~~~~~~~~~~~~~~~~~~~~~~~~~~~~~~~~~~~~~~~~

This auxiliary transport equation is solved once (not self-consistently) for each value
of $\lambda$ with $\whsig$ as a pre-determined input function. The diffuse boundary
condition for $\whg_\lambda$ is given by Eq. (\ref{eq:ovchcomm}) with $\whg_{\text{\tiny TDL}}$
fixed by the self-consitently determined solution of the quasiclassical equations
and boundary condition for $\lambda=1$.

A further simplification for $\Del\Omega$ is possible in the weak-coupling limit when the self-energy
is purely off-diagonal and given by the order parameter, $\whsig=\whDelta$. The
self-consistency equation,
%~~~~~~~~~~~~~~~~~~~~~~~~~~~~~~~~~~~~~~~~~~~~~~~~~~~~~~~~~~~~~~~~~~~~~~~~~~~~~~~~~~~~~~~~~~~~
\be
\whDelta=2\,{\delta \Del\Phi[\whg]\over\delta \whg^{\text{\tiny tr}}}\,,
\label{eq:a2-feD}
\ee
%~~~~~~~~~~~~~~~~~~~~~~~~~~~~~~~~~~~~~~~~~~~~~~~~~~~~~~~~~~~~~~~~~~~~~~~~~~~~~~~~~~~~~~~~~~~~
can be used to evaluate $\Del\Phi[\whg]=\onefourth\text{Sp}'\left\{\whDelta\,\whg\right\}$.
The resulting free energy reduces to
%~~~~~~~~~~~~~~~~~~~~~~~~~~~~~~~~~~~~~~~~~~~~~~~~~~~~~~~~~~~~~~~~~~~~~~~~~~~~~~~~~~~~~~~~~~~~
\be
\Del\Omega =
\onehalf\int_0^1\,d\lambda\,
\text{Sp}'\left\{\whDelta\,\left(\whg_\lambda-\onehalf\whg\right)\right\}
\,.
\label{eq:a2-FE}
\ee

%\bibliographystyle{apsrev}
%%\bibliography{QFS,CM,candidbib}
%\bibliography{$HOME/script/bibliography/QFS,$HOME/script/bibliography/CM,$HOME/script/bibliography/Books}

\end{document}